\documentclass[review,authoryear]{elsarticle}
\usepackage{graphicx,afterpage}
\usepackage{bm}
\usepackage{soul}
\usepackage{empheq}
\usepackage{mathtools}
\usepackage[margin=1.25in]{geometry}
\usepackage{mathrsfs,color}
\usepackage{amsfonts}\usepackage{multirow}
\usepackage{amssymb}\usepackage{caption,comment}
\usepackage{amsmath}\usepackage{float}
\usepackage{amssymb,amsthm}
\usepackage{graphicx,afterpage,cancel}
\usepackage{epstopdf}
\usepackage{natbib}

\usepackage{hyperref}
\hypersetup{CJKbookmarks,%
bookmarksnumbered,%
colorlinks,%
linkcolor=black,%
citecolor=black,%
plainpages=false,%
pdfstartview=FitH,
pdfauthor=author}
\newtheorem{theorem}{Theorem}[section]

\newtheorem{proposition}[theorem]{Proposition}

\definecolor{orange}{RGB}{255,127,0}

\def\d{{\, \rm d}}

\usepackage{soul}
\usepackage{epstopdf}
\graphicspath{ {./Figs/}}
\afterpage{\clearpage}
\begin{document}

\begin{frontmatter}

\title{Data-Driven Statistical Reduced-Order Modeling and Quantification of Polycrystal Mechanics Leading to Porosity-Based Ductile Damage}

\author[1]{Yinling Zhang}
\author[1]{Nan Chen\corref{cor1}}
\ead{chennan@math.wisc.edu}
\author[2]{Curt A. Bronkhorst}
\author[3]{Hansohl Cho}
\author[1]{Robert Argus}

\cortext[cor1]{Corresponding author}

\address[1]{Department of Mathematics, University of Wisconsin, Madison, WI 53706, USA}
\address[2]{Department of Mechanical Engineering, University of Wisconsin, Madison, WI 53706, USA}
\address[3]{Department of Aerospace Engineering, Korea Advanced Institute of Science and Technology, Daejeon, 34141, Republic of Korea}

%\author{Yinling Zhang\textsuperscript{a}, Nan Chen\textsuperscript{a*}, Curt A. Bronkhorst\textsuperscript{b}, Hansohl Cho\textsuperscript{c}, Robert Argus\textsuperscript{a}}

%\noindent \textsuperscript{a}Department of Mathematics, University of Wisconsin, Madison, WI 53706, USA\\
%\textsuperscript{b}Department of Mechanical Engineering, University of Wisconsin, Madison, WI 53706, USA\\
%\textsuperscript{c}Department of Aerospace Engineering, Korea Advanced Institute of Science and Technology, Daejeon, 34141, Republic of Korea\\
%\textsuperscript{*}Corresponding Authour: chennan@math.wisc.edu

\begin{abstract}
Predicting the process of porosity-based ductile damage in polycrystalline metallic materials is an essential practical topic. Ductile damage and its precursors are represented by extreme values in stress and material state quantities, the spatial probability density function (PDF) of which are highly non-Gaussian with strong fat tails. Traditional deterministic forecasts utilizing sophisticated continuum-based physical models generally lack in representing the statistics of structural evolution during material deformation. Computational tools which do represent complex structural evolution are typically expensive. The inevitable model error and the lack of uncertainty quantification may also induce significant forecast biases, especially in predicting the extreme events associated with ductile damage. In this paper, a data-driven statistical reduced-order modeling framework is developed to provide a probabilistic forecast of the deformation process of a polycrystal aggregate leading to porosity-based ductile damage with uncertainty quantification. The framework starts with computing the time evolution of the leading few moments of specific state variables from the spatiotemporal solution of full-field polycrystal simulations. Then a sparse model identification algorithm based on causation entropy, including essential physical constraints, is utilized to discover the governing equations of these moments. An approximate solution of the time evolution of the PDF is obtained from the predicted moments exploiting the maximum entropy principle. Numerical experiments based on polycrystal realizations of a representative body-centered cubic (BCC) tantalum illustrate a skillful reduced-order model in characterizing the time evolution of the non-Gaussian PDF of the von Mises stress and quantifying the probability of extreme events. The learning process also reveals that the mean stress is not simply an additive forcing to drive the higher-order moments and extreme events. Instead, it interacts with the latter in a strongly nonlinear and multiplicative fashion. In addition, the calibrated moment equations provide a reasonably accurate forecast when applied to the realizations outside the training data set, indicating the robustness of the model and the skill for extrapolation. Finally, an information-based measurement is employed to quantitatively justify that the leading four moments are sufficient to characterize the crucial highly non-Gaussian features throughout the entire deformation history considered.
\end{abstract}


\end{frontmatter}

\section{Introduction}

For high-pressure mechanical loading conditions of polycrystalline metallic materials with ready access to mobile dislocations, it is known that material structure is a critical factor in the formation and placement of cavitated pores \citep{WILKERSON201721, NGUYEN20191, PhysRevLett.117.215503, doi:10.1063/1.3607294, doi:10.1080/14786435.2012.734638, doi:10.1063/1.4941823}. The structure of metallic materials also evolves with deformation, so the material’s initial state and the physical processes invoked during deformation create a complex spatially dependent structural evolution problem. Certainly, environmental and boundary conditions are essential in how these materials resist deformation and choose physical pathways to accommodate these boundary conditions \citep{10.1007/978-1-4614-4238-7_44, doi:10.1063/1.371527, doi:10.1063/1.5045045}. Regarding material cavitation, pores are discrete localized features that form at very small length scales to contribute to a larger field of pores over much larger length scales. The formation of each pore is considered to be a rare event in response to the extreme mechanical loading of an aggregate composite polycrystalline metal. This produces a spatially varying mechanical response dependent upon the initial and evolved heterogeneous structural state of the material that then creates a time-evolving heterogeneous stress state within the material, which drives pore formation \citep{doi:10.1063/1.3607294, doi:10.1080/14786435.2012.734638}. The heterogeneous structural state of the material also contributes to spatial variation of defects which create regions of mechanical weakening that are more prone to cavitation. Although the qualitative picture which is painted is likely true and has been known for some time, we are lacking in our understanding of the important variables. Currently, we have not adequately quantified this statistical deformation process leading to these rare and discrete pore formation events \citep{LIEBERMAN2016270,doi:10.1080/02664763.2019.1686131}. The same heterogeneous conditions are also prevalent as the pores grow in size and ultimately percolate into a large-scale failed region.

The macroscale ductile damage models which currently exist to represent the formation and evolution of porosity fields under conditions of dynamic loading are continuum theories and employ largely mean-valued (first moment) material internal state variables to describe the evolution of stress, strain, material structure, and porosity with deformation over time \citep{doi:10.1063/1.371527, 10.1115/1.2899463, 10.1115/1.2895993, MOLINARI20011497, Czarnota2006, CZARNOTA20081624, doi:10.1063/1.4941823, VERSINO2018395, LONG2016611, WANG19942139, doi:10.1063/1.365320, doi:10.1063/1.357730}. Within these continuum structural field theories, the physics of finite elasticity and dislocation slip or deformation twinning mediated plasticity is most often represented by theories that assume an average over a finite-sized representative volume much larger than the mean size of a single crystal and certainly a nucleated pore. Although consistent with the spirit of these mean-valued continuum theories, the fundamental statistics of the structural evolution of the material with deformation leading to the rare events of pore formation are missing or only loosely considered. It is then perhaps of little surprise that we are still challenged in our ability to predict ductile damage in polycrystalline metallic materials exposed to extreme states of mechanical loading. These theories are generally meant to serve as structural design tools for components that can be large in size. Computational efficiency is often important; therefore, directly supplementing such theories with concurrent lower-length scale physics calculations that offer statistical insight is generally prohibitively expensive. Enriching conventional approaches to modeling ductile damage using continuum mean-valued field theories with a higher-order statistical representation of critical state variables, which is computationally tractable, is an important goal.

Structural features of a metallic material, which are defects that alter the resistance stress of pore formation, are varied and depend upon the type of material \citep{BIELER20091655, LIU2018438, REVILBAUDARD2016100}. Metallic alloys may have spatial variations in chemistry and phase \citep{FENG2003269,GU2023118715}. Commercially pure materials may have impurity sites that form inclusions or taint grain boundaries \citep{BENZERGA2010169, Gray9440, BUTLER2018248}. Each material may have several potential defect types, which have different spatial distributions and weakening influences on the lattice for pore formation. In principle, each defect type must be quantified for its impact on the material depending upon loading conditions. Several defect types may contribute to pore formation, or one may be dominant. High-purity materials will generally have fewer possible defect types and is therefore a logical choice to begin the process of quantifying influence. In the case of body-centered cubic (BCC) high-purity tantalum, it has been demonstrated that pores form predominantly at grain boundaries \citep{Gray_2014, bronkhorst2021local} for light to moderate shock loading conditions. Therefore, we seek to use polycrystal mechanics and physics tools to enable the quantification of grain size scale deformation behavior and statistics.

Dislocation motion in body-centered cubic (BCC) materials is generally observed to be dominated by the behavior of screw dislocations. These screw dislocations under conditions of zero velocity have been shown to dissociate into three partial dislocations, which form an equidistant triad within the crystallographic lattice \citep{PhysRevB.89.024104, Frederiksen2003365, Ismail-Beigi20001499, woodward2002flexible, GROGER20085401, Li2004104113, mrovec2004bond, PhysRevB.88.134101, Vitek2004415, VITEK200431, Xu19966941}. Since, at rest, this screw dislocation core is spread onto three different crystallographic planes, there is ambiguity in that it is no longer planar so that its subsequent motion cannot accurately be represented by a resolved shear stress based upon the classical Schmid rule \citep{GROGER20085401, VITEK200431, Vitek2004415, QIN1992835, ASARO19831, ASARO1977309, DAO1993143, CERECEDA2016242, KOESTER20123894, PATRA20141}. In addition, the spreading of the screw dislocation core also significantly reduces its mobility in relation to an edge dislocation. Therefore, the screw dislocation behavior dominates the plastic response of BCC materials. Although outside the scope of the present work, there is much we do not yet understand about dislocation motion in BCC metallic materials. The prevailing hypothesis that this core must be re-planarized to enable the nucleation of a kink band by one Burgers vector and the spreading of the kink band by the relative ease of motion of the formed edge dislocations at each of the two kinks \citep{BUTLER2018248}. The implication for this work is that BCC tantalum is represented by a model which accounts for the phenomenology of this screw dislocation behavior with the implication that its motion on a given slip system is directional and adds additional relevant and interesting heterogeneity.

Computational techniques to simulate the full-field response of polycrystal aggregates to mechanical and thermal loading have become important in the pursuit of structure-property-processing relationships for advancing our theoretical/computational tools and to speed the time-efficient design of new materials \citep{GhoshICME, DAO1993143, LEBENSOHN20136918, WHELAN2019106673, ARORA2020114, ZHANG2018258, SHANTHRAJ20131091, SHANTHRAJ2012154, Wu2016MicrostructuralMO}. It has been clearly demonstrated in the literature that the conditions of stress in these polycrystal aggregates are spatially variable with the characteristic length of the single crystal when the source of inelastic deformation is limited to dislocation motion and deformation twinning or structural phase transformation is missing \citep{BECKER20041983}. Such is the case for moderately shock-loaded tantalum \citep{bronkhorst2021local}. The disparity in the characteristic size of dislocations and grain size in these BCC materials affords the opportunity to represent the dislocation mediated plasticity in the grains comprising the aggregate by a continuum single crystal model with the appropriate deformation rate and temperature physics specific to BCC materials already discussed. Such calculations and their implication for material damage physics are the focus of the present work.

In this paper, a new data-driven reduced-order modeling framework is developed and applied to study the process of porosity-based ductile damage in polycrystalline metallic materials. Different from the traditional reduced-order models aiming to capture large-scale dynamical features using reduced-order bases, the framework developed here focuses on the time evolution of critical statistics that characterizes the intrinsic dynamical properties from a probabilistic viewpoint. In other words, the quantity of interest in such a statistical reduced-order model is the probability density function (PDF) associated with some crucial material state variables in the original physical system, for example, the von Mises stress. As ductile damage and its precursors appear as extreme values, the PDF of the von Mises stress and other state variables manifests strong non-Gaussian features \citep{bronkhorst2021local}. In particular, the PDF for the examples examined here often has a one-sided fat tail corresponding to the extreme stress values during the ductile damage process. Therefore, an appropriate characterization of the time evolution of such a non-Gaussian fat-tailed PDF advances the understanding of the ductile damage process of polycrystalline metallic materials in the statistical sense. It also offers the possibility to facilitate studying the formation of extreme damage events, including discovering the statistical precursor, the onset, and the occurrence of these damage events with proper uncertainty quantification represented by the tail probability. In addition, modeling the statistical evolution of important material state variables advances predicting the timing and probability of the occurrence of extreme events with a much lower computational cost than the sophisticated physical models generally used for sub-granular physics resolution.

The development of such a data-driven statistical reduced-order model here starts with the spatiotemporal solution of statistical volume element full-field polycrystal simulations using a computationally expensive physical single-crystal model. At each fixed time instant, the solution of the von Mises stress or other material state variables over different spatial grid points is utilized to compute the statistics. Note that, as the PDF is a function of both time and the values of the material state variables, its governing equation is given by a partial differential equation (PDE) \citep{kadanoff2000statistical,10.1115/1.4054237,BERDICHEVSKY2023105102}. However, simple and explicit expressions of such a PDE are often unavailable for such extremely complicated systems. Despite the recent development of many machine learning methods to discover this PDE \citep{xu2020solving, zhai2022deep, zhang2022physically, camporeale2022data}, the computational cost of direct numerical methods usually remains demanding. To this end, an alternative approach that focuses on the time evolution of only the leading few moments is adopted here. Since the underlying dynamics of each moment equation is driven by an ordinary differential equation (ODE), the learning process becomes much more computationally affordable. Based on the moments computed from such a low-order model, an approximate solution of the time evolution of the PDF can be found in light of an information criterion called the maximum entropy principle \citep{jaynes1957information, karmeshu2003entropy, majda2006nonlinear, branicki2013non}. Thus, the remaining focus is on identifying the governing equations of the moments. A simple learning algorithm via causation entropy  \citep{elinger2020information, elinger2021causation} and an explicit parameter estimation approach containing essential physical constraints \citep{chen2020learning, chen2022causality} are utilized to achieve this goal, which provides a sparse identification of the model structure.

The data-driven statistical reduced-order modeling framework developed here has several unique features. First, due to the inevitable model error and the lack of uncertainty quantification, the traditional deterministic forecast via sophisticated macroscale continuum ductile damage models may suffer from significant forecast biases, especially in predicting the extreme events associated with ductile damage that is often sensitive to the model uncertainty. Particularly, traditional continuum-based ductile damage models are based upon mean-value state variables and, therefore, cannot properly capture these extreme value events. In contrast, the statistical reduced-order model developed here provides a possible pathway to understand and represent the statistical process of ductile damage in polycrystalline metallic materials within macroscale model frameworks. As the extreme events of damage and its precursor depend on many unresolved or small-scale physical effects, a statistical characterization of their development and emergence with an appropriate uncertainty quantification via the forecast probability is a preferable approach, which also indicates the predictability of these events. Second, compared with solving the physical models, calculating the leading few moments associated with certain critical state variables is computationally more efficient. It allows for carrying out a vast number of statistical predictions for material damage under different loading conditions within a short allotted time. Third, information theory can be applied to systematically quantify error and uncertainty. In addition to reconstructing an approximate PDF from the moments utilizing the maximum entropy principle, another information criterion, called relative entropy \citep{kullback1951information, kullback1959statistics,kleeman2011information, majda2006nonlinear}, is a natural choice to assess the difference between the reconstructed PDF and the truth in an unbiased fashion. This provides a rigorous assessment of the intrinsic barrier in characterizing the statistics under such a reduced-order approximation. Notably, the relative entropy highlights the tail probability, making it a suitable metric to quantify the uncertainty in the statistical prediction of extreme events. Finally, the statistical reduced-order modeling framework developed here, including the identification procedure, is simple and is generalizable to other materials physics applications with strongly non-Gaussian features and extreme events.

The framework developed here has broad scientific and computational implications. One critical issue in studying and forecasting the ductile damage process is understanding how low-order statistical moments affect the dynamics of high-order moments. In general, computing the time evolution of the low-order moments, such as the mean, is straightforward in most computational materials mechanics settings regardless of characteristic length with relatively low cost. Such statistics are also relatively easy to measure from lab experiments since they only rely on a small number of samples or the diagnostics used naturally provide a mean response of the material. Discovering the relationship between low- and high-order moments significantly facilitates inferring the latter that is crucial to the uncertainty quantification of extreme damage events. Such a relationship can be used to advantage with highly homogenized theories of material mechanics and physics. This topic will be studied here as an application of the framework.

The reminder of the paper is organized as follows. Section \ref{Sec:Framework} presents the general data-driven statistical reduced-order modeling framework. Section \ref{Sec:Polycrystal_Model} introduces the polycrystal model, which is utilized to create the spatiotemporal data for the development of the statistical reduced-order model. Section \ref{Sec:Results} shows the results of the statistical quantification of dynamical loading from the reduced-order model. Conclusion and discussions are included in Section \ref{Sec:Conclusion}.

\section{Data-Driven Statistical Reduced-Order Modeling Framework}\label{Sec:Framework}
\begin{figure}[ht]\centering
\hspace*{-0cm}\includegraphics[width=1.0\textwidth]{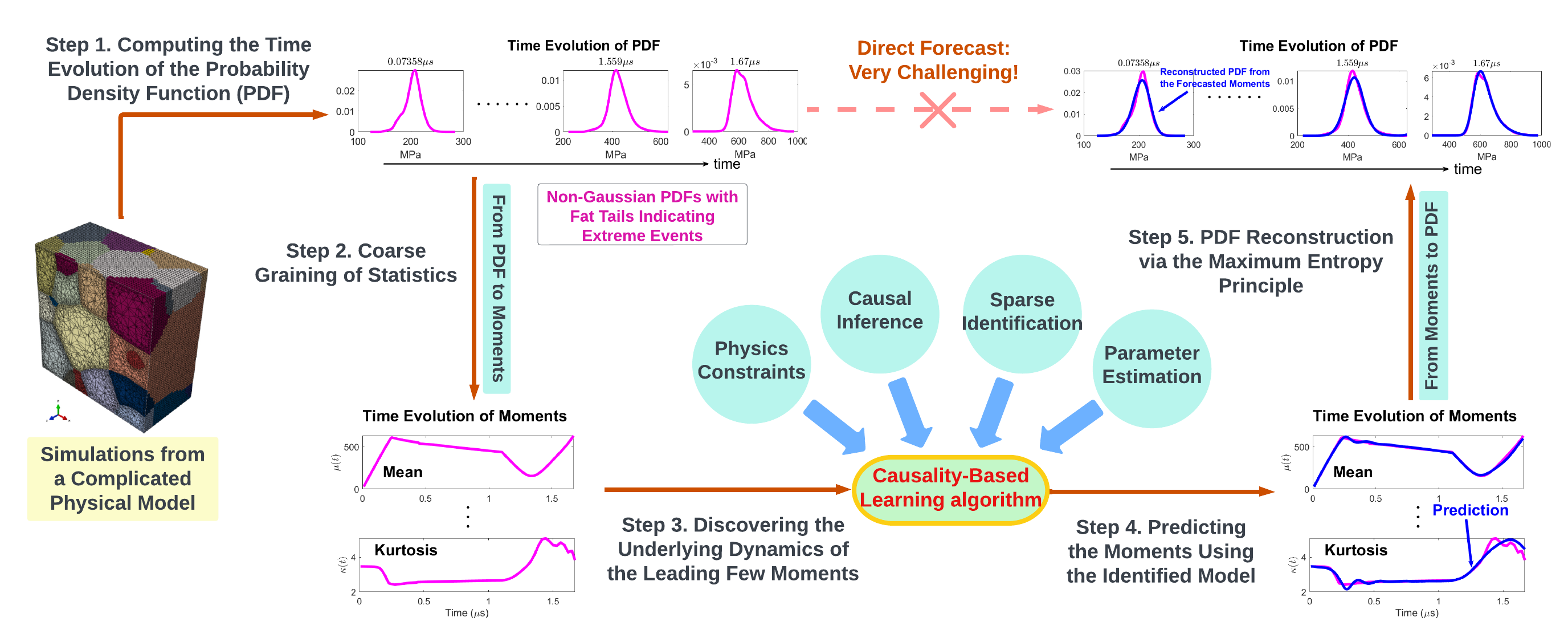}
\caption{Overview of the data-driven statistical reduced-order modeling framework.   }\label{Overview_Illustration}
\end{figure}

After obtaining the spatiotemporal solution describing the process of deformation leading to porosity-based ductile damage from a polycrystal simulation utilizing a sophisticated physical model, the data-driven statistical reduced-order modeling framework is as follows.
\begin{enumerate}
  \item Compute the time evolution of the PDF associated with a certain state variable from the spatiotemporal solution.
  \item Compute the time evolution of the leading few moments of the state variable to reach a coarse-grained representation of the statistics.
  \item Discover the underlying dynamics of these moments.
  \item Apply the identified model to forecast the moments.
  \item Reconstruct the time evolution of the PDF using the predicted moments via the maximum entropy principle.
\end{enumerate}
Figure \ref{Overview_Illustration} provides a schematic diagram containing an overview of the above procedure. The single crystal theory and computational polycrystal models that create the polycrystal simulation results will be presented in Section \ref{Sec:Polycrystal_Model}. The remaining section contains the details of the statistical reduced-order modeling procedure.

Denote by $u$ a scalar state variable of the statistical reduced-order model, for example, the von Mises stress.

\subsection{Computing the PDF and the moments}
At each time instant, denote by $u(x_i,y_j,z_k)$ the value of $u$ at the spatial location $(x_i,y_j,z_k)$ under the three-dimensional coordinate system, where $i=1,\ldots,I$, $j=1,\ldots,J$ and $k=1,\ldots,K$. The PDF and the moments are computed based on the total $IJK$ data points. The PDF $p(u)$ is given by a normalized histogram that satisfies the condition $\int_{-\infty}^\infty p(u) \d u= 1$. The $n$-th centralized moment is defined as \citep{gardiner2004handbook, gupta2020fundamentals}
\begin{equation}\label{Definition_Moments}
\begin{split}
  \mu_1 &= \int_{-\infty}^\infty u p(u) \d u,\qquad \mbox{for~} n = 1,\\
  \mu_n &= \int_{-\infty}^\infty (u-\mu_1)^n p(u) \d u,\qquad \mbox{for~} n\geq2.
\end{split}
\end{equation}
After normalizing the moments for $n\geq 3$ by taking $\tilde{\mu}_n=\mu_n/(\mu_2)^{n/2}$, each of the scaled moments $\tilde{\mu}_n$ describes a particular aspect of the corresponding PDF. Further define $\tilde{\mu}_1=\mu_1$ and $\tilde{\mu}_2=\mu_2$. The leading few scaled moments $\tilde{\mu}_n$ with $n=1,\ldots,N$ are used in developing the statistical reduced-order model as the coarse-grained representation of the statistics. Hereafter $\tilde{\mu}_n$ is called the $n$-th moment for the conciseness of presentation. To allow the reconstruction of a non-Gaussian PDF via the maximum entropy principle, the following two conditions need to be satisfied \citep{majda2006nonlinear}. First, at least $N\geq3$ moments are required in the PDF reconstruction to include the information beyond the Gaussian statistics. Second, $N$ must be an even number; otherwise the maximum entropy principle fails to provide a well-defined PDF. Therefore, the minimum number of the scaled moments to be adopted is $N=4$. The leading four scaled moments are mean, variance, skewness, and kurtosis. Skewness is a measure of the asymmetry of a PDF. For a unimodal distribution, negative skew commonly indicates that the tail is on the left side of the distribution, and positive skew indicates that the tail is on the right. Kurtosis describes the tail behavior of a PDF. Higher kurtosis corresponds to a larger probability of extreme events or outliers. The kurtosis for a Gaussian distribution has a value of $3$. When the kurtosis exceeds $3$, the distribution usually has fat tails which may correspond to the triggering of extreme events. For problems involving material damage, the shape of the PDF and tail character becomes very important.

In the following, $N=4$, namely the leading four scaled moments, are utilized in developing the statistical reduced-order model for the process of the ductile damage in polycrystal simulations. Such a choice of $N$ will be justified by computing the lack of information in the reconstructed PDF based on these moments and the truth.

\subsection{Discovery of the moment equations via a causality-based sparse identification learning algorithm}
\begin{figure}[ht]\centering
\hspace*{-0cm}\includegraphics[width=0.6\textwidth]{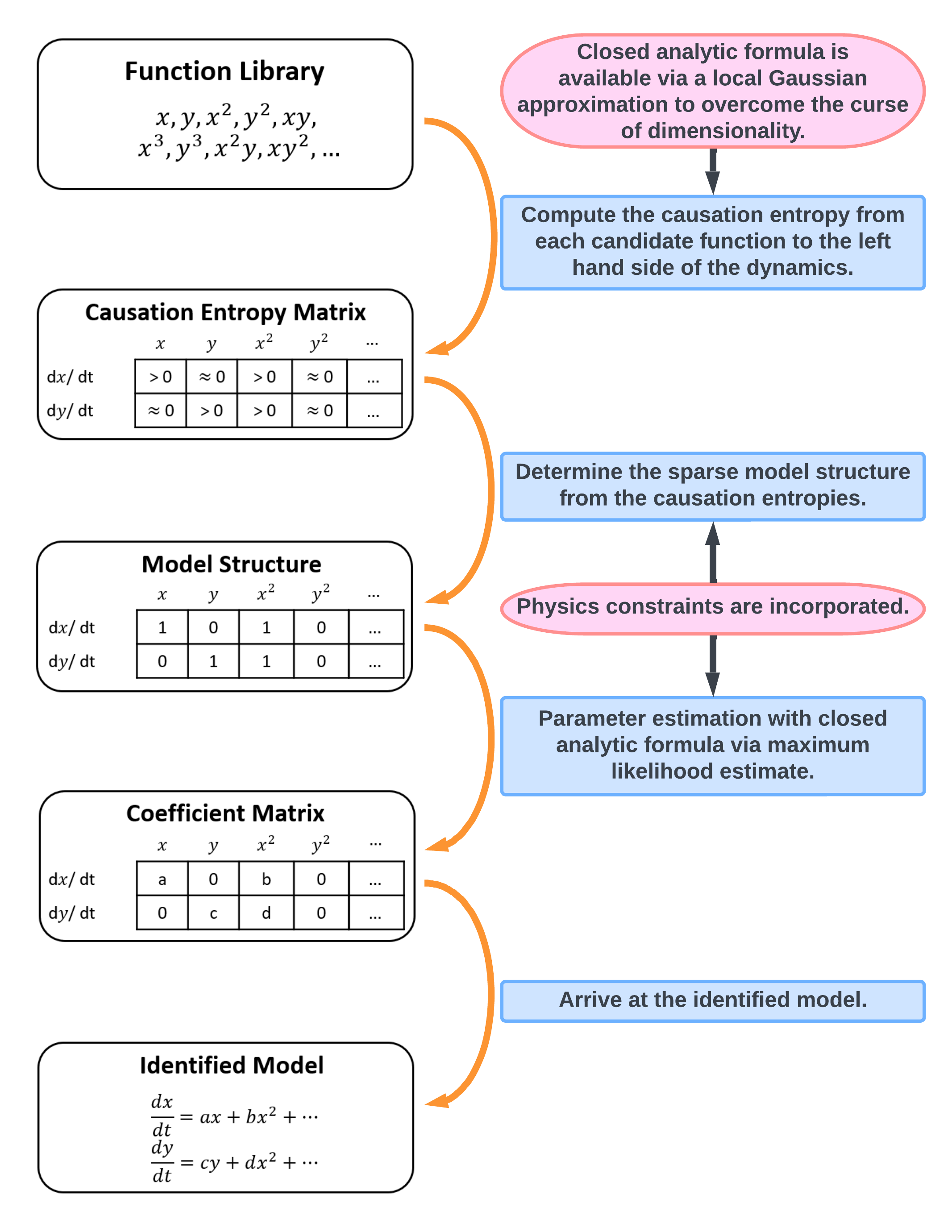}
\caption{Schematic illustration of the causality-based learning algorithm of underlying dynamics.  }\label{CEM_Illustration}
\end{figure}

To discover the possible underlying dynamics of the leading $N$ moments, a causality-based learning algorithm is utilized \citep{elinger2020information, elinger2021causation, chen2022causality}. The method is summarized in Figure \ref{CEM_Illustration} with the following steps.

\subsubsection{Developing a function library}
First, a library $\mathbf{f}$ consisting of in total $M$ possible candidate functions to model the right-hand side of the moment equations is proposed,
\begin{equation}\label{Library}
  \mathbf{f} = \{f_1,\ldots, f_{m-1}, f_m, f_{m+1}, \ldots, f_M\}.
\end{equation}
The development of such a library is subjective based on empirical and expert knowledge. Typically, a large number of candidate functions is included in the library to allow for coverage over different possible dynamical features of the underlying unknown true system. This also allows for the opportunity to learn new physical relationships from the physical information.

\subsubsection{Computing the causation entropy}
Next, a causation entropy $C_{f_{m} \rightarrow \dot{\tilde{\mu}}_n \mid\left[\mathbf{f} \backslash {f}_{m}\right]}$ is computed to detect if the candidate function $f_m$ contributes to the dynamics of $\tilde{\mu}_n$, namely $\d \tilde{\mu}_n/\d t :=\dot{\tilde{\mu}}_n$. It reads \citep{elinger2021causation}:
\begin{equation}\label{Causation_Entropy}
  C_{f_{m} \rightarrow \dot{\tilde{\mu}}_n \mid\left[\mathbf{f} \backslash {f}_{m}\right]}=H(\dot{\tilde{\mu}}_n|\left[\mathbf{f} \backslash {f}_{m}\right]) - H(\dot{\tilde{\mu}}_n|\mathbf{f}).
\end{equation}
In Eq. \eqref{Causation_Entropy}, the set $\mathbf{f} \backslash {f}_{m}$ represent a new set that includes all functions in $\mathbf{f}$ except $f_m$. It thus contains $M-1$ candidate functions and is defined as
\begin{equation}\label{Library2}
  \mathbf{f}\backslash {f}_{m} = \{f_1,\ldots, f_{m-1},f_{m+1},\ldots, f_M\}.
\end{equation}
The term $H(\cdot|\cdot)$ is the conditional entropy, which is related to Shannon's entropy $H(\cdot)$ and the joint entropy $H(\cdot,
\cdot)$. For two multi-dimensional state variables $\mathbf{X}$ and $\mathbf{Y}$, they are defined as \citep{cover1999elements}:
\begin{equation}\label{Entropies}
\begin{split}
  H(\mathbf{X}) &= -\int_x p(\mathbf{x})\log(p(\mathbf{x}))\d \mathbf{x},\\
  H(\mathbf{Y}| \mathbf{X}) &= -\int_\mathbf{x}\int_\mathbf{y} p(\mathbf{x},\mathbf{y})\log(p(\mathbf{y}|\mathbf{x}))\d \mathbf{y}\d \mathbf{x},\\
  H(\mathbf{X},\mathbf{Y}) &= -\int_\mathbf{x}\int_\mathbf{y} p(\mathbf{x},\mathbf{y})\log(p(\mathbf{x},\mathbf{y}))\d \mathbf{y}\d \mathbf{x}.
\end{split}
\end{equation}
On the right-hand side of Eq. \eqref{Causation_Entropy}, the difference between the two conditional entropies indicates the information in $\dot{\tilde{\mu}}_n$ contributed by the specific function $f_m$ given the contributions from all the other functions. Thus, it tells if $f_m$ provides additional information to $\dot{\tilde{\mu}}_n$ conditioned on the other potential terms in the dynamics. It is worthwhile to highlight that the causation entropy in Eq. \eqref{Causation_Entropy} is fundamentally different from directly computing the correlation between $\dot{\tilde{\mu}}_n$ and $f_m$, as the causation entropy also considers the influence of the other library functions. If both $\dot{\tilde{\mu}}_n$ and $f_m$ are caused by a common factor $f_{m^\prime}$, then $\dot{\tilde{\mu}}_n$ and $f_m$  can be highly correlated. Yet, in such a case, the causation entropy $C_{f_{m} \rightarrow \dot{\tilde{\mu}}_n \mid\left[\mathbf{f} \backslash {f}_{m}\right]}$ will be zero as $f_m$ is not the causation of $\dot{\tilde{\mu}}_n$.

The causation entropy is computed from each of the candidate functions in $\mathbf{f}$ to each $\dot{\tilde{\mu}}_n$. Thus, there are in total $NM$ causation entropies, which can be written as a $N\times M$ matrix, called the causation entropy matrix.

Note that the dimension $\mathbf{X}$ in Eq. \eqref{Entropies} is $M$ when it is applied to compute the second term on the right-hand side of the causation entropy in Eq. \eqref{Causation_Entropy}. This implies that the direct calculation of the entropies in Eq. \eqref{Entropies} involves a high-dimensional numerical integration, which is, however, a well-known computationally challenging issue \citep{bellman1961dynamic}. To circumvent the direct numerical integration, the entropy calculation is based on the following approximate method that utilizes a Gaussian fit to each distribution.
\begin{proposition}[Computational approximation of the causation entropy]
By approximating all the joint and marginal distributions as Gaussians, the  causation entropy can be computed in the following way:
\begin{equation}
\label{Entropy_Gaussians}
\begin{split}
C_{\mathbf{Z} \rightarrow \mathbf{X} | \mathbf{Y}} &=H(\mathbf{X} | \mathbf{Y})-H(\mathbf{X} | \mathbf{Y}, \mathbf{Z}) \\
& = H(\mathbf{X},\mathbf{Y}) - H(\mathbf{Y}) - H(\mathbf{X},\mathbf{Y},\mathbf{Z}) + H(\mathbf{Y},\mathbf{Z})\\
& = \frac{1}{2} \ln(\operatorname{det}(\mathbf{R}_{\mathbf{X}\mathbf{Y}}))-\frac{1}{2} \ln(\operatorname{det}(\mathbf{R}_{\mathbf{Y}})) - \frac{1}{2} \ln(\operatorname{det}(\mathbf{R}_{\mathbf{X}\mathbf{Y}\mathbf{Z}})) \\
& \quad +\frac{1}{2} \ln(\operatorname{det}(\mathbf{R}_{\mathbf{Y}\mathbf{Z}})),
\end{split}
\end{equation}
where $\mathbf{R}_{\mathbf{XYZ}}$ denotes the covariance matrix of the state variables $(\mathbf{X},\mathbf{Y},\mathbf{Z})$ and similar for other covariances. The notations $\ln(\cdot)$ and $\operatorname{det}(\cdot)$ are the logarithm of a number and determinant of a matrix, respectively.
\end{proposition}
The explicit expression in Eq. \eqref{Entropy_Gaussians} based on the Gaussian approximation can efficiently compute the causation entropy. It allows the computation of the causation entropy with a moderately large dimension, which is typically the case for a low-order system containing the leading few moment equations. It is worth noting that the Gaussian approximation may lead to certain errors in computing the causation entropy if the true distribution is highly non-Gaussian. Nevertheless, the primary goal here is not to obtain the exact value of the causation entropy. Instead, it suffices to detect if the causation entropy $C_{f_{m} \rightarrow \dot{\tilde{\mu}}_n \mid\left[\mathbf{f} \backslash {f}_{m}\right]}$ is nonzero (or practically above a small threshold value). In most applications, if a significant causal relationship is detected in the higher-order moments, it is very likely in the Gaussian approximation. This allows us to efficiently determine the sparse model structure, where the exact values of the nonzero coefficients on the right-hand side of the identified model will be calculated via a simple maximum likelihood estimation to be discussed in the following. Note that the Gaussian approximation has been widely applied to compute various information measurements and leads to reasonably accurate results \citep{majda2018model, tippett2004measuring, kleeman2011information, branicki2012quantifying}.

\subsubsection{Determining the model structure with sparse identification}
With the $N\times M$ causation entropy matrix in hand, the next step is determining the model structure. This can be done by first setting up a threshold value of the causation entropy and then retaining only those candidate functions that have the causation entropies exceeding the threshold. This allows the identified system to contain only functions that have significant contributions to the dynamics. In other words, the identified system has a sparse model structure. Sparsity is crucial to discovering the correct underlying physics and prevents overfitting \citep{ying2019overview, brunton2016discovering}. It will also guarantee the robustness of the model in response to perturbations and allow the model to apply to certain extrapolation tests, for example, the statistical forecast of other data sets with similar but not identical features.

Physical or statistical constraints can be added to the above model identification procedure. One important constraint is called the physics constraint. It means the conservation of energy in the quadratic nonlinear terms are important properties in many complex turbulent systems \citep{majda2012physics, harlim2014ensemble}. The physics constraint prevents the finite-time blow-up of the solutions and facilitates a skillful medium- to long-range forecast. To incorporate the physics constraint into the model identification, it is essential to keep or remove the related terms simultaneously.

\subsubsection{Parameter estimation}\label{Subsec:ParameterEstimation}
The final step is to estimate the parameters in the identified model.
Denote by $\mathbf{U}$ an $N$-dimensional column vector containing the leading $N$ moments, namely $\mathbf{U}=(\tilde{\mu}_1,\ldots, \tilde{\mu}_N)^\mathtt{T}$, where $\cdot^{\mathtt{T}}$ is the vector transpose. The identified system consisting of the moment equations can be written in the vector form:
\begin{equation}\label{Identified_Model}
  \frac{\d \mathbf{U}}{\d t} = \boldsymbol\Phi(\mathbf{U}),
\end{equation}
where $\boldsymbol\Phi(\mathbf{U})$ is an $N$-dimensional column vector with each entry representing the summation of the remaining nonlinear candidate functions from the causality-based learning algorithm for each component of $\mathbf{U}$. Denote by $\boldsymbol\Theta\in \mathbb{R}^s$ a column vector containing all the parameters in the identified model with $s\gg n$ in a typical situation. The right-hand side of Eq. \eqref{Identified_Model} can be written as
\begin{equation}\label{RHS_Model}
  \boldsymbol\Phi(\mathbf{U}) = \mathbf{M} \boldsymbol\Theta + \mathbf{Q},
\end{equation}
where $\mathbf{M}\in\mathbb{R}^{N\times s}$ is a matrix each entry of which is a function of $\mathbf{U}$ while $\mathbf{Q}\in \mathbb{R}^N$ is a column vector that depends on $\mathbf{U}$. Only $\mathbf{M}$ is multiplied by the parameters $\boldsymbol\Theta$.

Assume data comes in a sufficiently high frequency in time, which is the typical case if the data is computed from a complicated physical model and the frequency of data dumps with time is completely controlled. Then consider a temporal discretization of Eq. \eqref{Identified_Model} using a first-order forward Euler scheme from time $t_j$ to $t_{j+1}$ \citep{gardiner2004handbook}. Further assign the data value to the variables at both time instants. This leads to the following relationship from $t_j$ to $t_{j+1}$,
\begin{equation}\label{discrete_eqn}
  \mathbf{U}^{j+1} = \mathbf{U}^j + \widetilde{\boldsymbol\Phi}(\mathbf{U}^j) \Delta{t} + \boldsymbol\sigma\boldsymbol\varepsilon^j\sqrt{\Delta{t}},
\end{equation}
where $\Delta{t}$ is a small time step from $t_j$ to $t_{j+1}$, and $\boldsymbol\varepsilon^j$ is an independent and identically distributed (i.i.d.) standard multidimensional Gaussian random number, representing the residual of the parameter fitting, and $\boldsymbol\sigma\in\mathbb{R}^{N\times N}$ is a diagonal matrix representing the strength of the residual. Further assume $\boldsymbol\sigma$ is a constant matrix over time.

Due to the Euler approximation, the one-step time evolution from $\mathbf{U}^j$ to $\mathbf{U}^{j+1}$ is approximated by a linear function within such a short time interval. Therefore, the likelihood can be computed based on a Gaussian distribution,
\begin{equation}
  \mathcal{N} (\boldsymbol\mu^j, \boldsymbol\Sigma) = C |\boldsymbol\Sigma|^{-\frac{1}{2}} \exp \left(- \frac{1}{2} (\mathbf{U}^{j+1} - \boldsymbol\mu^j)^\mathtt{T} (\boldsymbol\Sigma)^{-1} (\mathbf{U}^{j+1} - \boldsymbol\mu^j)\right),
\end{equation}
where $C$ is a normalization constant. The mean and the covariance of such a Gaussian distribution are given by $\boldsymbol\mu^j = \mathbf{M}^j \boldsymbol\Theta + \mathbf{Q}^j+ \mathbf{U}^j$ and $\boldsymbol\Sigma = \boldsymbol\sigma\boldsymbol\sigma^\mathtt{T}\Delta{t}$, respectively. Taking the logarithm to cancel the exponential function and summing up the likelihood over the entire time period yield
\begin{equation}\label{eq: SI_obj}
  \mathcal{L} =  \frac{1}{2} \sum_{j=1}^J  (\mathbf{U}^{j+1} - \mathbf{M}^j \boldsymbol\Theta - \mathbf{Q}^j - \mathbf{U}^j)^\mathtt{T}  (\boldsymbol\Sigma)^{-1} (\mathbf{U}^{j+1} - \mathbf{M}^j \boldsymbol\Theta - \mathbf{Q}^j - \mathbf{U}^j)  - \frac{J}{2} \log |\boldsymbol\Sigma| ,
\end{equation}
where $J+1=\lfloor T/\Delta{t}\rfloor$ with $\lfloor\cdot\rfloor$ being the floor function that rounds down the result to the nearest integer.
To find the minimum of~$\mathcal{L}$, it is sufficient to compute the zeros of $\frac{\partial\mathcal{L}}{\partial \boldsymbol\Theta} =0$ and $\frac{\partial \mathcal{L}}{\partial \boldsymbol\Sigma} =0$, which leads to
\begin{subequations}\label{eq: SI_Equation_R_Theta}
    \begin{align}
        \boldsymbol\Sigma &= \frac{1}{J} \sum_{j=1}^J (\mathbf{U}^{j+1} - \mathbf{M}^j \boldsymbol\Theta - \mathbf{Q}^j-\mathbf{U}^j)(\mathbf{U}^{j+1} - \mathbf{M}^j\boldsymbol\Theta - \mathbf{Q}^j-\mathbf{U}^j)^\mathtt{T} , \label{eq: SI_Equation_R_Theta1}\\
        \boldsymbol\Theta &= \mathbf{D}^{-1} \mathbf{c},\label{eq: SI_Equation_R_Theta2}
    \end{align}
\end{subequations}
where
\begin{equation}\label{eq: SI_aux_physics_constraint}
	\mathbf{D} = \sum_{j=1}^J (\mathbf{M}^j)^\mathtt{T} \boldsymbol{\Sigma}^{-1}\mathbf{M}^j \quad \text{and} \quad \mathbf{c} = \sum_{j=1}^J(\mathbf{M}^j)^\mathtt{T} \boldsymbol{\Sigma}^{-1}(\mathbf{U}^{j+1} - \mathbf{Q}^j-\mathbf{U}^j).
\end{equation}
The equations in Eq. \eqref{eq: SI_Equation_R_Theta} are solved by first setting $\boldsymbol\Theta=\mathbf{0}$ to find $\boldsymbol\Sigma$ in Eq. \eqref{eq: SI_Equation_R_Theta1} via essentially the quadratic variation, and then plugging in the result into Eq. \eqref{eq: SI_Equation_R_Theta2} and Eq. \eqref{eq: SI_aux_physics_constraint} to obtain $\boldsymbol\Theta$. The analytic solution in Eq. \eqref{eq: SI_Equation_R_Theta}--Eq. \eqref{eq: SI_aux_physics_constraint} significantly facilitates the estimation of the parameter values compared with using a numerical solver for a non-convex optimization problem resulting from a constrained regression based on, for example, an L1 penalty \citep{santosa1986linear, tibshirani1996regression}. The computational cost in Eq. \eqref{eq: SI_Equation_R_Theta}--Eq. \eqref{eq: SI_aux_physics_constraint} is proportional to the square of the number of functions in the identified model (in computing $\mathbf{D}$) and to the total number of observational points in time. Since the goal is to find a parsimonious model, the number of the candidate functions remaining in the model is expected to be small, which leads to a relatively low computational cost in solving Eq. \eqref{eq: SI_Equation_R_Theta}--Eq. \eqref{eq: SI_aux_physics_constraint}.

In addition to the information provided by the data, including prior knowledge, such as the physical laws, can significantly facilitate the learning process. Such additional knowledge often appears in the form of constraints, many of which are characterized by imposing certain relationships in the model parameters.
In such a situation, the above formulae can be slightly adjusted. Assume the constraints are imposed to the model parameters with the following form,
\begin{equation}
	\mathbf{H} \boldsymbol\Theta = \mathbf{g},
\end{equation}
where $\mathbf{H}$ and $\mathbf{g}$ are constant matrices. To incorporate these constraints, the Lagrangian multiplier method is applied, which modifies the  objective function in Eq. \eqref{eq: SI_obj},
\begin{equation}\label{eq: objective_constraint}
\begin{gathered}
	\mathcal{L} = \frac{1}{2} \sum_{j=1}^J (\mathbf{U}^{j+1} - \mathbf{M}^j \boldsymbol\Theta - \mathbf{Q}^j- \mathbf{U}^j)^\mathtt{T}  (\boldsymbol\Sigma)^{-1} (\mathbf{U}^{j+1} - \mathbf{M}^j \boldsymbol\Theta - \mathbf{Q}^j - \mathbf{U}^j)\\ - \frac{J}{2}\log |\boldsymbol\Sigma^{-1}|  + \boldsymbol{\lambda}^\mathtt{T}  (\mathbf{H} \boldsymbol\Theta - \mathbf{g}),
\end{gathered}
\end{equation}
where $\boldsymbol\lambda$ is the Lagrangian multiplier.
The solution to the minimization problem with the new objective function Eq. \eqref{eq: objective_constraint} is given as follows \citep{boyd2004convex},
\begin{subequations}\label{eq: SI_Equation_R_Theta_lambda}
    \begin{align}
        \boldsymbol\Sigma &= \frac{1}{J} \sum_{j=1}^J  (\mathbf{U}^{j+1} - \mathbf{M}^j \boldsymbol\Theta - \mathbf{Q}^j - \mathbf{U}^j)(\mathbf{U}^{j+1} - \mathbf{M}^j\boldsymbol\Theta - \mathbf{Q}^j- \mathbf{U}^j)^\mathtt{T} \label{eq: SI_Equation_R_Theta_lambda1}\\
        \boldsymbol\lambda &= \left(\mathbf{H} \mathbf{D}^{-1} \mathbf{H}^\mathtt{T} \right)^{-1} (\mathbf{H} \mathbf{D}^{-1}\mathbf{c} - \mathbf{g}), \\
        \boldsymbol\Theta &= \mathbf{D}^{-1} \left( \mathbf{c} - \mathbf{H}^\mathtt{T} \boldsymbol\lambda \right),
    \end{align}
\end{subequations}
where $\mathbf{D}$ and $\mathbf{c}$ are defined in~Eq. \eqref{eq: SI_aux_physics_constraint}.

\subsection{PDF reconstruction via maximum entropy principle}
The identified model from the above learning algorithm gives the value $g_n$ of the moment $\tilde{\mu}_n$ for $n=1,\ldots,N$ at each time instant. These $N$ moments can be utilized to reconstruct a least biased approximate PDF via the so-called maximum entropy principle \citep{jaynes1957information, karmeshu2003entropy}.
\begin{proposition}\label{Proposition:Maximum_Entropy_Principle}
Define the set $\mathcal{C}_N$ that contains all possible distributions, the leading $N$ moments $\tilde{\mu}_n$ of which satisfy $\tilde{\mu}_n=g_n$ for $n=1,\ldots,N$:
\begin{equation}\label{Constraints}
    \mathcal{C}_N = \left\{p|\tilde{\mu}_n = g_n,~1\leq n\leq N\right\}.
\end{equation}
 The least biased probability distribution $\tilde{p}$ given the constraints $\mathcal{C}_N$ is the one that maximizes the entropy,
 \begin{equation}\label{Max_Entropy_Principle}
    \max_{p\in\mathcal{C}_N}\mathcal{S}(p) = \mathcal{S}(\tilde{p}),\qquad \tilde{p}\in\mathcal{C}_N,
\end{equation}
where entropy is defined as
  \begin{equation}\label{Shannon_Entropy_Continuous}
    \mathcal{S}(p) = -\int p(u) \ln(p(u)) \d u.
  \end{equation}
\end{proposition}
Therefore, solving the least biased distribution $\tilde{p}$ becomes a constrained optimization problem. The cost function is the entropy in Eq. \eqref{Max_Entropy_Principle}. The constraints are given by the $N$ equalities in the set Eq. \eqref{Constraints} together with the two additional constraints that all the distributions need to satisfy, that is, $\int p(u)\d u = 1$ and $p(u)\geq0$.

\subsection{Quantifying the lack of information in the reconstructed PDF via relative entropy}
With the reconstructed PDF in hand, it is essential to assess its difference (or the so-called lack of information) compared with the true distribution. The comparison involves three PDFs. They are (a) the true PDF computed from the solution of the physical model, (b) the reconstructed PDF using the leading few moments of the truth, and (c) the reconstructed PDF using the same number of moments as (b) but from running the moment equations. The difference between (b) and (a) characterizes the lack of information using the coarse-grained description of the statistics, namely what is lacking in the leading $N$ moments. Such a difference reflects the intrinsic barrier that cannot be overcome by further optimizing the reconstructed PDF unless an additional number of moments is included. A smaller difference is preferred as it implies a more skillful approximation using these $N$ moments. The difference between (c) and (b) accesses the model error of the moment equations. Such an error is expected to be minimized by the optimization procedure in the model identification and parameter estimation. The difference between (c) and (a) is the total error in the results from the moment equations compared with the truth.

The traditional path-wise measurements, such as the root-mean-square error or the standard Euclidean norms, are not appropriate to quantify the difference between the two distributions. The main reason is that these measurements give a very low weight toward the tails of the distributions, which are, however, the crucial part of describing the probability of extreme events. Therefore, an information criterion is introduced here, which is called the relative entropy, that can be utilized to quantify the difference between two PDFs $p$ and $\tilde{p}$, where $p$ is the truth or the one that contains more information while $\tilde{p}$ is the approximate PDF or the PDF which has an error. The relative entropy is given by \citep{majda2010quantifying, kleeman2011information}
\begin{equation}\label{Relative_Entropy}
    \mathcal{P}(p,\tilde{p}) = \int p(u) \ln\left(\frac{p(u)}{\tilde{p}(u)}\right) \d u.
\end{equation}
The relative entropy $\mathcal{P}(p,\tilde{p})$, which is also known as Kullback-Leibler (KL) divergence or information divergence \citep{kullback1997information, kullback1951information}, is an objective metric for the difference between two PDFs that measures the expected increase in ignorance, or lack of information, about the system incurred by using $p^M$, when the outcomes are generated by $p$. It has two important features. First, $\mathcal{P}(p,\tilde{p})$ is positive unless $p = \tilde{p}$. The relative entropy increases monotonically with the difference between the two PDFs. Second, $\mathcal{P}(p,\tilde{p})$ is invariant under general nonlinear change of variables. This is crucial in practice as different communities often use distinct coordinates and units. It also means that relative entropy is a dimensionless measurement. Finally, it is worthwhile to highlight that the ratio between $p(u)$ and $\tilde{p}(u)$ is utilized to characterize the difference between the two PDFs. This indicates that even though both the PDFs can have small values at the tail, the ratio $p(u)/\tilde{p}(u)$ can be large if the result from the moment equation (namely, $\tilde{p}(u)$) severely underestimates the probability of extreme events. In such a case, the relative entropy becomes large, indicating the lack of skill in $\tilde{p}(u)$ in quantifying the uncertainty of extreme events.

\subsection{Additional remarks}
The model identification method utilized here is only one of many sparse identification approaches \citep{brunton2016discovering, liu2021sparse, ibanez2018multidimensional, goyal2021learning, boninsegna2018sparse, cortiella2021sparse}. In principle, other model identification methods or computational algorithms can also be applied for the experiments in Section \ref{Sec:Results}. The reason for using this specific method is that it separates model identification from parameter estimation, where the latter can be efficiently solved with closed analytic formulae. For more complicated cases, additional latent processes may become essential to advance the model identification of the quantity of interest. In such a situation, an iterative method, which alternates between model identification, learning the latent processes, and parameter estimation, can be developed based on the framework utilized here \citep{chen2022causality}.

It is also worthwhile to notice that the variance and kurtosis are non-negative. This condition is satisfied in all the results presented in Section \ref{Sec:Results} since the values of these moments are typically far from zero, especially when extreme events are about to occur. A model creating negative variance and kurtosis is thus very unskillful and is unlikely to be selected by the learning algorithm. Yet, such a positivity condition is not automatically guaranteed within the current framework. A straightforward remedy to ensure the positivity of these moments is to identify the governing equation of the square root of the even-order moments and then take the square of the model simulation results to recover the moments.

\section{Soft-Coupled Damage and Polycrystal Models}\label{Sec:Polycrystal_Model}
%{\color{red}{Curt, I leave this section to you. Feel free to change the plan below. }}
%\begin{itemize}
%  \item A quick overview of the polycrystal model. There is no need to show the equations but some necessary high-level description of the model is needed.
%  \item Some key features of the models.
%  \item Show a model simulation. Include one or two figures, something like Fig 9 and Figs 14-15 (only stress) in \citep{bronkhorst2021local}. Add some discussions.
%\end{itemize}

\begin{figure}[ht]\centering
\hspace*{-0cm}\includegraphics[width=0.6\textwidth]{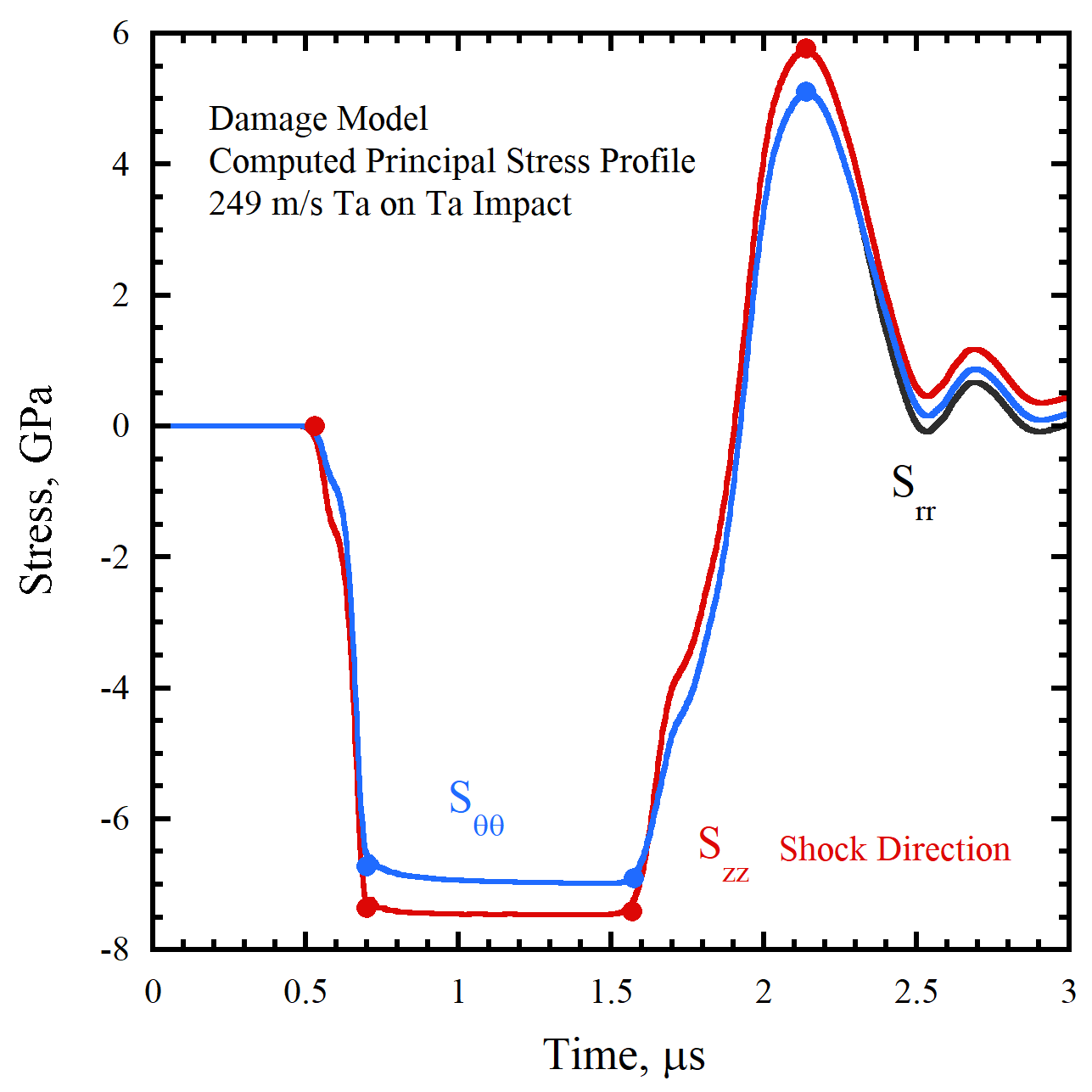}
\caption{Principal stress profile drawn from numerical simulation of a tantalum on tantalum plate impact experiment \citep{Gray_2014} using the damage model of \citep{bronkhorst2021local}. The points on the continuous stress profiles indicate end-points to a piece-wise linear approximation applied to the polycrystal models. }\label{Stress_time_profile}
\end{figure}

We seek a micromechanical understanding of the conditions which contribute to the formation of pores for states of high triaxiality loading and with a particular focus on a computational database for a high-purity tantalum material. The basis for this computational database of interest to this work is a plate impact experiment which is previously reported upon \citep{Gray_2014} with particular interest to the tantalum on tantalum experiment and material pedigree information. We employ both free surface velocity trace, and the cross-sectioned soft recovered sample, which is also examined in prior work \citep{doi:10.1063/1.4941823, bronkhorst2021local}. In \citep{bronkhorst2021local}, a macroscale continuum damage model from the pedigree \citep{10.1115/1.2899463, WANG19942139, doi:10.1063/1.365320, doi:10.1063/1.357730, doi:10.1063/1.371527, Czarnota2006, CZARNOTA20081624, MOLINARI20011497, 10.1115/1.2895993, VERSINO2018395} is developed and used to describe the experimental results described in \citep{Gray_2014}. The tantalum on tantalum plate impact experimental result, in particular, was well described, and the simulation results were found to be independent of computational element size by natural regularization of deformation rate sensitivity contained in the plasticity model and an inertial term which is included in the equation of motion of the thick-walled sphere representation of a pore. The experiment was material and geometry symmetric so that the center of the damage field in the shock direction was at the center of the sample. With the experimental free surface velocity trace and the damage field well represented by the numerical results of the damage model, these computed results were then used to quantify the principal stress history at the center of the sample to the point in time when the damage model indicated porosity initiation. This occurred at 2.140 $\mu$s after impact, as indicated by the time of maximum shock direction tensile stress. See Figure \ref{Stress_time_profile}. From this history of the principal stresses, one can see an initial period of rapid ramped increase in compressive stress as the shock wave reached the center of the sample. This is followed by a period of relatively constant stress over a period of 0.88 $\mu$s followed by rapid unloading and tensile reloading as the reflected tensile waves interact and lead to the onset of damage. This principal stress profile is approximated by three linear segments as indicated by the data points in Figure \ref{Stress_time_profile}. This piece-wise linear profile is then applied to a series of computational statistical volume elements (SVEs) representing the tantalum material used in the experimental study \citep{bronkhorst2021local}. Note that the profile used is offset in time by 0.47 $\mu$s, the time it takes the shock wave to travel to the center of the sample. The applied profile is given in Table \ref{Table:TimeProfile}.

\begin{table}[H]
\centering
\small
\begin{tabular}{|c|c|c|}
\hline
Time, $\mu$s & 1st Principal Stress, GPa & 2nd and 3rd Principal Stresses, GPa\\
\hline
0.000 & 0.000 & 0.000 \\
\hline
 0.230 & -7.370 & -6.730 \\
\hline
1.105 & -7.420 & -6.911 \\	
\hline
1.670 & 5.770 & 5.110 \\
\hline
\end{tabular}
\caption{Time profile of the magnitude of principal stresses extracted from the center location of the tantalum on tantalum plate impact simulation to the time when porosity nucleation began. Note that the time for the shock wave to travel to the center location (0.470 $\mu$s) was subtracted from each time value relative to the profile of Figure \ref{Stress_time_profile}.}\label{Table:TimeProfile}
\end{table}

\begin{figure}[ht]\centering
\hspace*{-0cm}\includegraphics[width=1.0\textwidth]{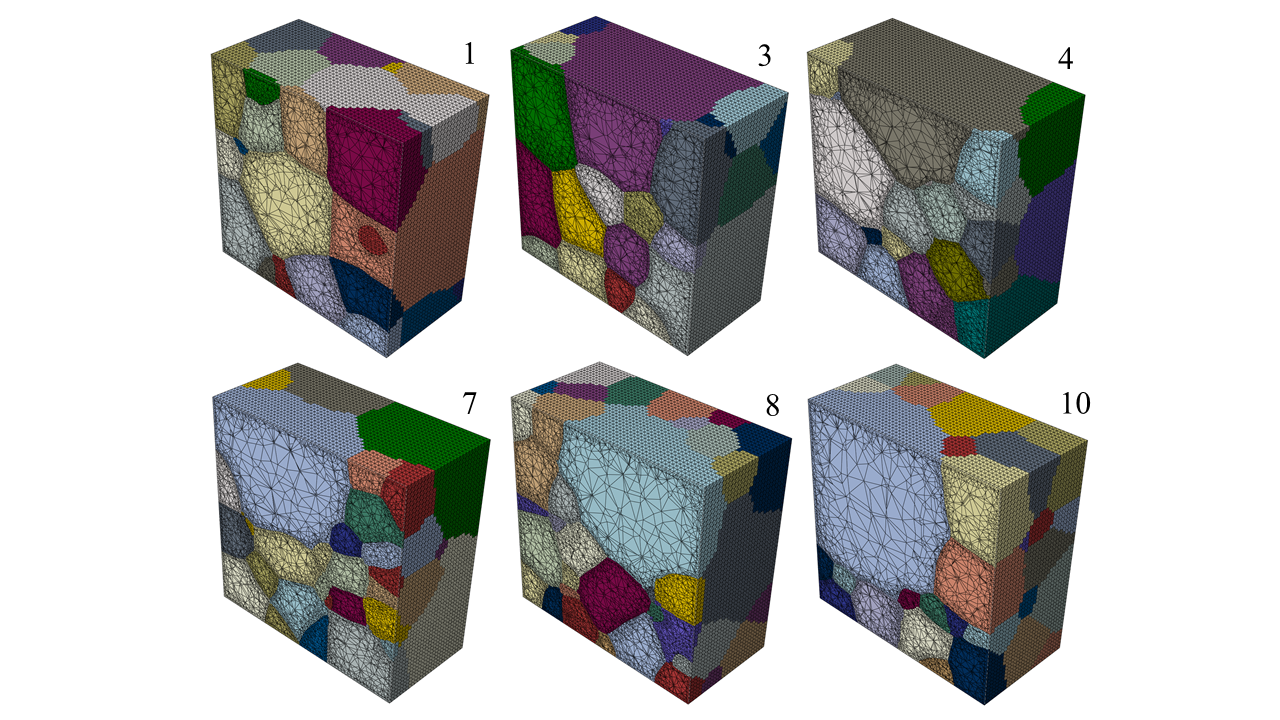}
\caption{ Cross-sectional images of six of ten statistical volume element polycrystal models used for this work showing the original geometry and crystal structure. Realization 1 model is utilized as the training data for the statistical reduced-order model in Section \ref{Sec:Results} while the other five are utilized for the forecast (extrapolation) experiments. }\label{Poly_Meshes}
\end{figure}

\begin{figure}
    \centering
    \hspace*{-0cm}\includegraphics[width=1.0\textwidth]{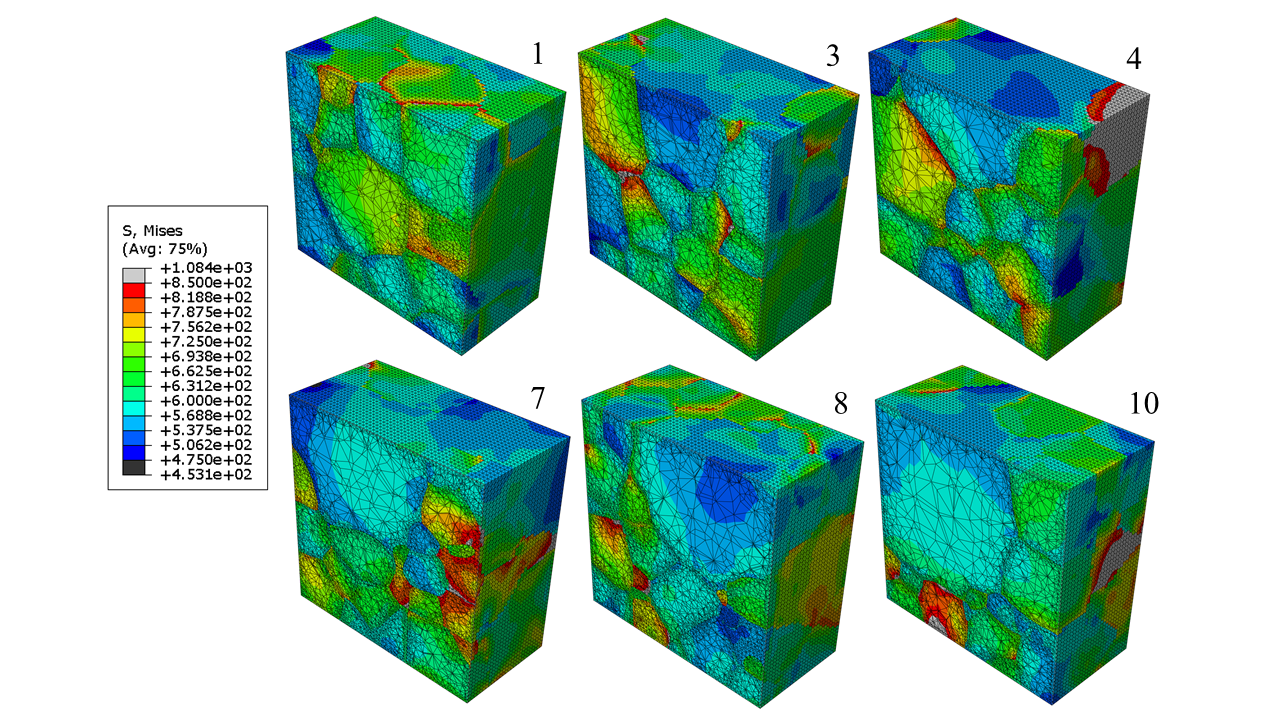}
    \caption{Cross-sectional images of von Mises stress (MPa) in the six SVE simulations at the maximum tensile stress state and the condition at which the damage model suggested porosity nucleation began.}
    \label{Contours}
\end{figure}

The loading profile derived from the computed representation of experimental data using the damage model \citep{VERSINO2018395,bronkhorst2021local} and defined in Table \ref{Table:TimeProfile} is applied to a series of ten SVE polycrystal models of the experimental material. These models were constructed based upon the tools and methodology reported by \citep{KNEZEVIC2014239}. The term SVE differs from that of representative volume element (RVE) in that the volume of sampling of the former may not be large enough to adequately represent the statistical response of larger volumes. For example, two SVE units will provide differing stress response to deformation and the PDF of state variables of interest, while two RVE units will provide similar measured response. The tantalum material microstructure was characterized by electron-backscatter diffraction (EBSD) metallography of the three principal plate directions. These EBSD data sets were then used within the open-source code Dream.3D \citep{GROEBER20081274} to construct statistically representative 3D numerical microstructure cubes with between 65 and 100 grains represented within each volume. The methodology discussed in \citep{KNEZEVIC2014239,KNEZEVIC201493} uses STL files produced by Dream.3D to construct tetrahedral mesh tessellations of each grain and renders grain boundary conforming polycrystal models. The crystallographic orientations assigned to each grain in the ten SVEs were chosen by Dream3D to match the experimental EBSD crystallographic texture. This is important in our study of damage in polycrystalline materials. With each cube face constrained to remain flat and unrotated, each of the ten polycrystal realizations was loaded by applying the stress profile given in Table \ref{Table:TimeProfile} to the primary faces of each cube. Cross-sectional images of six of these polycrystal models are shown in Figure \ref{Poly_Meshes}. Cross-sectional images of von Mises stress contour plots in the six SVE simulations at the maximum tensile stress state and the condition at which the damage model suggested porosity nucleation began are illustrated in Figure \ref{Contours}.

The single crystal theory used to describe the deformation behavior of the tantalum material was developed along the lines of prior work by \citep{ASARO1977309, ACHARYA20002213,KOTHARI199851, BUSSO19961, bronkhorst1992polycrystalline, KALIDINDI1992537, ANAND20045359, BRONKHORST20072351, GURTIN2000989, gurtin2010mechanics, ALLEMAN2015176, CHO2018138, bronkhorst2021local}. The single crystal theory and results are to be found in \citep{bronkhorst2021local} and so will only be briefly summarized here for completeness.

The model is written for large deformation and therefore the total deformation gradient is multiplicatively decomposed in elastic (with crystallographic rotation) and plastic components by $\mathbf{F}=\mathbf{F}^{*}\mathbf{F}^{p}$. Accordingly for crystal mechanics, the plastic velocity gradient is given by

\begin{equation}\label{Lp}
    \mathbf{L}^{p}=\sum_{\alpha=1}^N \dot{\gamma}^{\alpha}_{p} \mathbf{S}^{\alpha}_{0},
\end{equation}
where $\dot{\gamma}^{\alpha}_{p}$ is the plastic shear rate on slip system $\alpha$ and $\mathbf{S}^{\alpha}_{0}$ is the Schmid projection tensor for slip system $\alpha$. The second Piola-Kirchhoff stress is given as a function of the energy conjugate Green-Lagrange elastic strain and the thermal expansion term
\begin{equation}\label{2PK}
\mathbf{T}^{*} = \mathbb{C}(\theta) [ \mathbf{E}^{*}-\mathbf{A} \left( \theta-\theta_0 \right) ],
\end{equation}
for temperature $\theta$, reference temperature $\theta_{0}$, thermal expansion coefficient tensor $\mathbf{A}$, and elastic stiffness tensor $\mathbb{C}(\theta)$. The elastic Green-Lagrange strain is defined by
\begin{equation}\label{GLstrain}
    \mathbf{E}^{*}=\frac{1}{2} \left( {\mathbf{F}^{*}}^{T} \mathbf{F}^{*} - \mathbf{1} \right).
\end{equation}
For interests of this work, dislocation motion is well within the regime of thermally activated glide \citep{kocks1975thermodynamics} and therefore the flow rule is given by
\begin{equation}\label{flowrule}
\dot{\gamma}^{\alpha}_{p}=\dot{\gamma}^{0}_{p} \text{exp} \left( -\frac{\Delta G}{k_{B}\theta} \left\langle 1 - \left\langle \frac{\tau_{eff}^{\alpha}}{\Tilde{s}^{\alpha}_{l}} \right\rangle ^{p} \right\rangle ^{q} \right),\text{    for } \tau^{\alpha}>0,\text{ otherwise }\dot{\gamma}_{p}=0,
\end{equation}
where $\dot{\gamma}_0$ is the reference shear rate, $\Delta G$ is activation energy, $\tau^{\alpha}_{eff}=\tilde{\tau}^{\alpha}-\tilde{s}^{\alpha}$ is the non-Schmid resolved effective shear stress for slip system $\alpha$, $\tilde{s}^{\alpha}$ is the resistance to dislocation motion due to dislocation interactions, and $\tilde{s}^{\alpha}_{l}$ is the intrinsic lattice resistance. The brackets $\langle \cdot \rangle=\frac{1}{2}(|\cdot|+(\cdot))$. The modified resolved shear stress including the non-glide stresses for BCC crystal systems is $\tilde{\tau}^{\alpha}=\mathbf{T}^{*} : (\mathbf{S}^{\alpha}_{0}+\tilde{\mathbf{S}}^{\alpha}_{0})$. The Schmid tensor $\mathbf{S}^{\alpha}_{0}$ and the additional projection tensor involving non-glide stresses $\tilde{\mathbf{S}}^{\alpha}_{0}$ are given by
\begin{subequations}\label{NStensors}
\begin{align}
\mathbf{S}^{\alpha}_{0}&=\mathbf{m}^{\alpha}_{0}\otimes\mathbf{n}^{\alpha}_{0}, \\
\tilde{\mathbf{S}}^{\alpha}_{0}&=\omega_{1}\mathbf{m}^{\alpha}_{0}\otimes\mathbf{n}^{\prime\alpha}_{0}+\omega_2\left(\mathbf{n}^{\alpha}_{0}\times\mathbf{m}^{\alpha}_{0}\right)\otimes\mathbf{n}^{\alpha}_{0}+\omega_{3}\left(\mathbf{n}^{\prime\alpha}_{0}\times\mathbf{m}^{\alpha}_{0}\right)\otimes\mathbf{n}^{\prime\alpha}_{0}, \\
\omega_{i}&=\omega_{i,ss}+\left(\omega_{i,0K}-\omega_{i,ss}\right)\text{exp}\left(-\theta/\theta_{r}\right),
\end{align}
\end{subequations}
where $\mathbf{m}^{\alpha}_{0}$ and $\mathbf{n}^{\alpha}_{0}$ are slip system $\alpha$ direction and plane normal respectively, $\mathbf{n}^{\prime\alpha}_{0}$ is an additional non-Schmid direction, and $\omega_{i}$ are empirical weighting factors. The dislocation structure and intrinsic lattice resistances are given by
\begin{equation}\label{resist}
\tilde{s}^{\alpha}=s^{\alpha}\frac{\mu(\theta)}{\mu_{0}}, \hspace{1cm}
\tilde{s}^{\alpha}_{l}=s^{\alpha}_{l}\frac{\mu(\theta)}{\mu_{0}},
\end{equation}
where $s^{\alpha}$ and $s^{\alpha}_{l}$ are the structural resistance and the intrinsic lattice resistance at 0 K, respectively, $\mu_{0}$ is the anisotropic shear modulus at 0 K and $\mu(\theta)$ is the temperature-dependent anisotropic shear modulus at current temperature. The rate and temperature dependent saturation value of the structural resistance is given by
\begin{equation}
    s^{\alpha}_{ss}=s^{\alpha}_{ss,0}\left(\frac{\dot{\gamma}^{\alpha}_{p}}{\dot{\gamma}^{0}_{p}}\right)^\frac{k\theta}{A},
\end{equation}
where $s^{\alpha}_{ss,0}$ is reference resistance and $A$ is reference energy. The structural resistance evolves with dislocation structure evolution during plastic deformation and is given by
\begin{subequations}\label{sevolve}
\begin{align}
\dot{s}^{\alpha}&=\sum_{\beta=1}^N h_{\alpha\beta}\left|\dot{\gamma}^{\beta}_{p}\right|, \\
h_{\alpha\beta}&=\left(q_{i}+\left(1+q_{i}\right)\delta_{\alpha\beta}\right)h_{\beta}, \\
h_{\beta}&=h_{0}\left(\frac{s^{\beta}_{ss}-s^{\beta}}{s^{\beta}_{ss}-s^{\beta}_{0}}\right),
\end{align}
\end{subequations}
with $q_{i}$ the latent hardening ratio, initial self-hardeing rate $h_{0}$, and the initial slip system resistance $s_{0}^{\beta}$. The full model, values of material parameters, the procedure by which they were evaluated, and the model performance compared against experimental results for Eq. \eqref{Lp}--Eq. \eqref{sevolve} can be found in \citep{bronkhorst2021local}.

\section{Statistical Quantification and Prediction of the Polycrystal Simulations}\label{Sec:Results}
\subsection{Setup}\label{Subsec:Setup}

The dynamic behavior of materials under shock loading conditions on polycrystal realizations of a representative BCC tantalum is studied here in reference to plate impact experiments. These are conceptually simple in that a stationary circular disk (the target sample) is impacted by another circular disk (the flyer) moving at high velocity. The flyer is accelerated through a gun barrel by either compressed gas or in some cases, ignited gunpowder. The flyer is soft-mounted on a sabot that travels through the barrel.

One simulation from the polycrystal model (the first one in Figure \ref{Poly_Meshes}; defined as simulation \#1 hereafter) is utilized to determine the moment equations. The cubic SVE has an initial dimension of $165 \mu m$ on a side. The simulation contains 69 grains with grain shape and crystallographic orientation populations derived from electron backscatter diffraction (EBSD) results and used in SVE construction by Dream3D. At the time $t=0$, principal stresses, as described earlier, are imposed on the faces of the cube with the time profile defined in Figure \ref{Stress_time_profile} and Table \ref{Table:TimeProfile}.
The entire simulation covers the period from $t=0$ to $t=1.67\mu s$, ending when the damage model indicates that porosity initiation occurs. Computed data is written for 48 sequential times during the loading profile for each of the six simulations considered in this work. At each time instant, there are $540,436$ data points in space corresponding to the number of integration points in the realization {\#1} model, sufficient to eliminate sampling error in computing the one-dimensional statistics. Although any material state variable can be characterized within this framework, the quantity of interest for the statistical reduced-order model studied here is the von Mises stress. The entire process can be divided into three stages. The first stage goes from $t=0$ to $t=0.23\mu s$, which is the period of compression ramp. The second stage goes from $t=0.23\mu s$ to $t=1.105\mu s$, representing the time of shock wave travel through the sample center with relatively constant compressive stress. The third stage is from $1.105\mu s$ onward, which is the period at the end of the compressive wave travel, unloading, and subsequent release wave interaction caused tensile loading leading to porosity nucleation.

Once the moment equations are determined, they will first be applied to predict the statistics for the same polycrystal simulation starting from $t=0$ to validate the learning results. Then, the model will be used to forecast the time evolution of the statistics for the remaining five polycrystal simulations  in Figure \ref{Poly_Meshes} with the same loading condition but with different numbers of grains and orientations. This is called an extrapolation test, which is crucial to demonstrate that the identified model is generalizable to different scenarios.

Two statistical model experiments are carried out here. In the first experiment, the data from the polycrystal simulation is utilized to learn a four-dimensional model that characterizes the time evolution of the leading four moments. In the second experiment, the time evolution of the mean is treated as known, and the goal is to identify the moment equations for the variance, skewness, and kurtosis, which is a three-dimensional system. The motivation of the second experiment is that the mean is usually computed by continuum-based damage models or measured accurately from experiments. However, the higher-order moments are typically not readily available. Therefore, finding a relationship between the mean and the higher-order moments has practical significance in characterizing and predicting the information associated with the extreme damage events contained in the higher-order moments. Discovering such a relationship also helps understand if the mean stress interacts with high-order moments and extreme events in a simple additive way or a more complicated nonlinear and multiplicative way.

\subsection{Non-Gaussian statistics and information characterized by the leading four moments}

\begin{figure}[ht]\centering
\hspace*{-0cm}\includegraphics[width=1.0\textwidth]{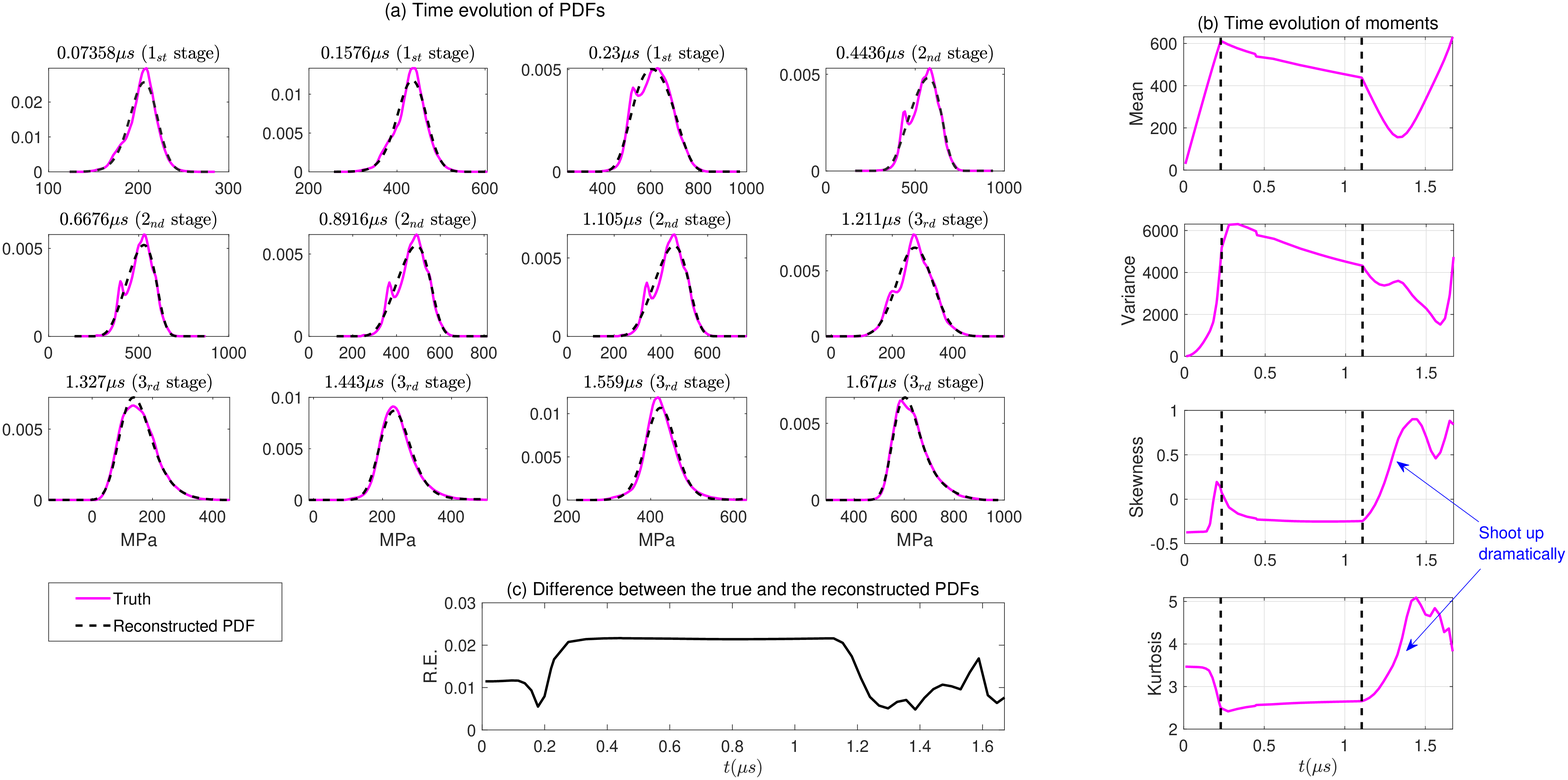}
\caption{True statistics of the von Mises stress computed from the polycrystal simulation \#1. Panel (a): time evolution of the PDFs (pink) and the reconstructed PDFs (black) using the leading four moments via the maximum entropy principle. Panel (b): time evolution of the leading four moments. Panel (c): time evolution of the relative entropy (R.E.) that quantifies the difference between the truth and the reconstructed PDF.}\label{Truth}
\end{figure}

Figure \ref{Truth} shows the statistics of the von Mises stress computed from the polycrystal simulation \#1. The time evolution of the statistics (pink color) demonstrate an apparent three-stage behavior consistent with the three-stage loading process described in Section \ref{Subsec:Setup}. During the first stage (up to $t=0.23\mu s$), the mean and the variance increase monotonically, which is a natural outcome due to the loading condition causing the material compression. The skewness remains near zero; therefore, the PDFs are symmetric around the mean. The kurtosis at the initial period is about $3.5$, which results in moderate fat tails, as is expected from the compression. At $t=0.23\mu s$, the kurtosis becomes less than $3$, suppressing the fat tails and leading to a sub-Gaussian distribution. Next, during the second stage (from $t=0.23\mu s$ to $t=1.105\mu s$), the sustained compressive stress with the shock wave traveling through the sample center induces a slow decaying of the mean and the variance. In contrast, the skewness and the kurtosis are nearly static. The skewness stays around a constant value of $0.2$ while the kurtosis is about $2.5$. The corresponding PDFs have weak bimodal structures. Yet, no significant extreme events are observed within this period due to the low kurtosis. These statistics experience the most significant changes during the third loading period (from $t=1.105\mu s$ to $t=1.67\mu s$). In addition to a first decrease and then an increase of the mean and the variance, both the skewness and the kurtosis increase dramatically. It is important to note that even though the stress profile applied to the SVE cube becomes zero while transitioning from compressive to tensile loading, the significant plastic deformation heterogeneity developed during the compressive loading process within the SVE maintains a value of von Mises stress of near 200 MPa at its minimum value. As a result, the PDF at each instant within this period manifests strong non-Gaussian features with a skewed profile and a one-sided fat tail towards the positive side that corresponds to extreme stress conditions, which may in turn cause damage events.

One crucial feature observed from Figure \ref{Truth} is that the reconstructed PDF using the leading four moments via the maximum entropy principle (black) successfully captures the true PDF (pink), especially the non-Gaussian characteristics. Despite small errors observed at the first few instants, the strongly skewed PDFs with fat tails are recovered accurately at later instants. As shown in Panel (c), the relative entropy that quantifies the gap between these two PDFs remains very low. It thus provides a quantitative justification that using the leading four moments is sufficient to characterize the time evolution of the dynamic loading process, including representing the statistical features of the development of extreme stress-driven events. Notably, the reconstructed PDFs accurately recover the tail probabilities when non-Gaussian fat-tailed distributions appear, implying that the uncertainty associated with the extreme events can be precisely characterized by the leading four moments.

\begin{figure}[ht]\centering
\hspace*{-0cm}\includegraphics[width=1.0\textwidth]{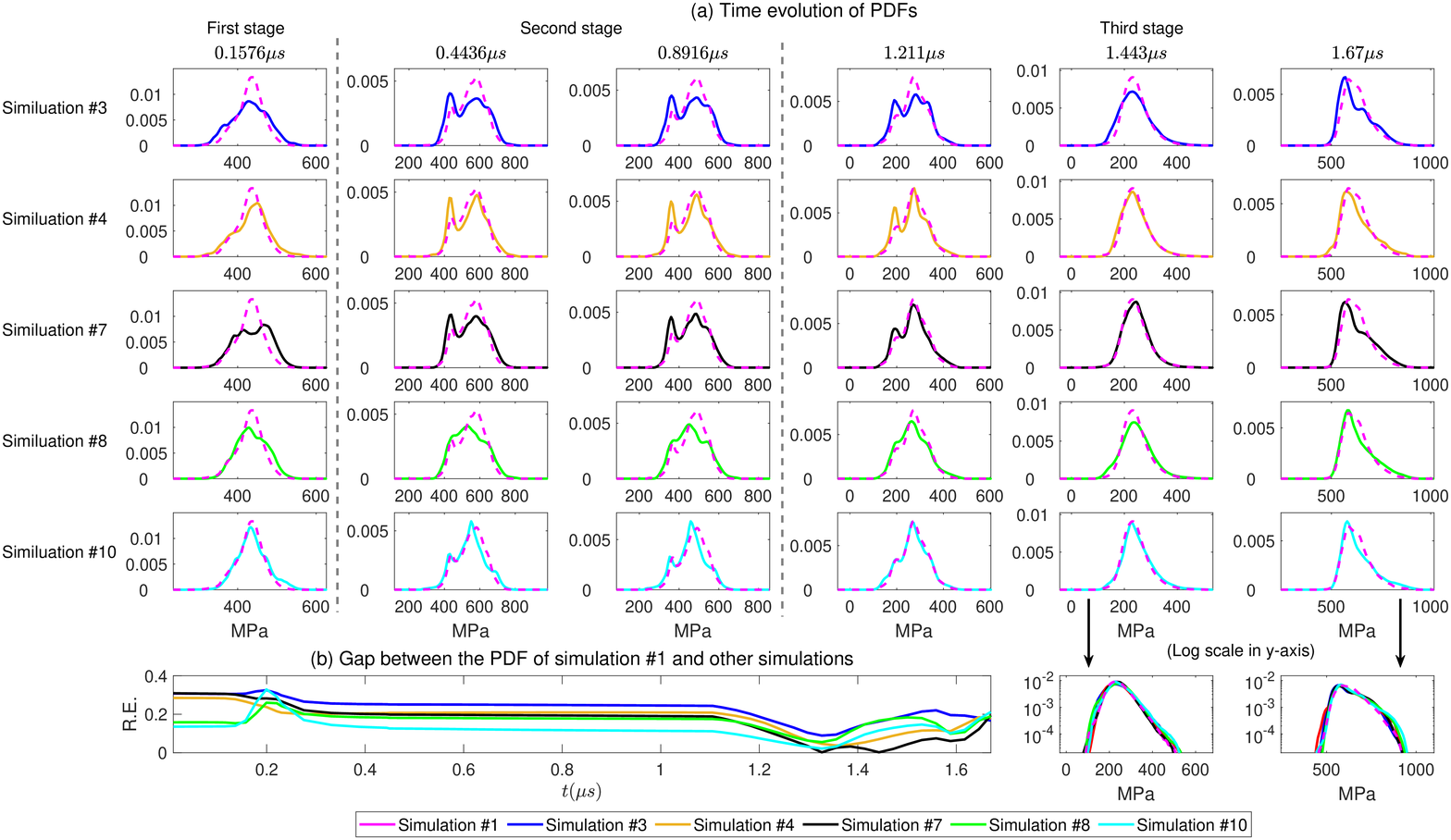}
\caption{Comparison of the training realization (simulation \#1) and the other five polycrystal realizations (simulation \#3, simulation \#4, simulation \#7, simulation \#8, and simulation \#10) that are utilized for the extrapolation tests. Panel (a): time evolution of the PDFs. Panel (b): time evolution of the relative entropy (R.E.) that quantifies the difference between the training realization and the other five crystal realizations. }\label{RE_othertests}
\end{figure}

Figure \ref{RE_othertests} includes a comparison of the PDFs between simulation \#1 and the other five simulations at different time instants. According to Panel (a), these simulations have overall similar time evolution of non-Gaussian statistics, especially the fat-tailed distributions at the final time instant. Yet, discrepancies are also seen between different realizations. Panel (b) shows the relative entropy as a function of time, which quantifies the difference between these realizations. Such differences provide a suitable case for the extrapolation tests. These results also clearly demonstrate the number of grains and total volume of the SVEs is not large enough to achieve statistical steady-state pertaining to the von Mises stress.

\subsection{Experiment I: Learning the moment equations of the leading four moments}
\begin{figure}[H]\centering
\hspace*{-0cm}\includegraphics[width=1.0\textwidth]{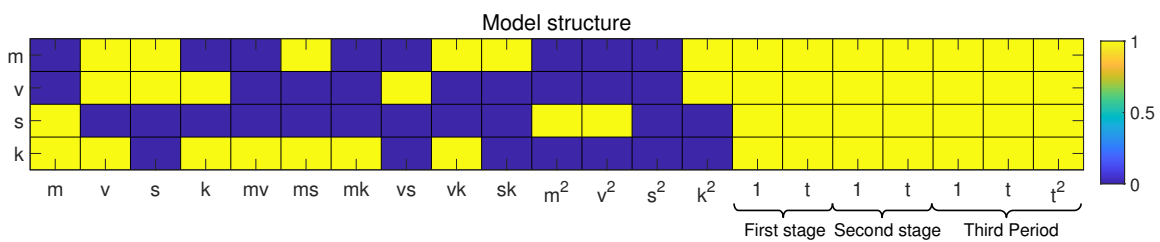}
\caption{The identified model structure of the statistical reduced-order model in Experiment I. The four rows show the identified structure on the right-hand side of the equation of mean ($m$), variance ($v$), skewness ($s$), and kurtosis ($k$), respectively. The terms represented by the yellow color are the ones that exist in the identified model while those indicated by the blue color do not appear in the model. The last seven columns show the structure of the three-stage forcing.}%{\color{red}{Yinling, replace this figure by the correct one.}}  }
\label{Model_Strucuture_Case1}
\end{figure}

In the first experiment, the data from the polycrystal simulation \#1 is utilized to learn the moment equations associated with the leading four moments. The library of candidate functions includes all the linear and quadratic nonlinear interactions between mean ($m$), variance ($v$), skewness ($s$), and kurtosis ($k$). The so-called physics constraint, introduced in Section \ref{Subsec:ParameterEstimation}, is incorporated into the model identification process. The physics constraint is one of the most fundamental constraints in modeling nonlinear dynamical systems \citep{majda2012physics, harlim2014ensemble, chen2018conditional}, which requires the total energy of the identified model resulting from the quadratic nonlinear terms to be conserved. This is motivated from the physical laws and energy exchange of many physical systems. Without taking into account the physics constraint, the identified models can have finite time blowup of solutions  and pathological behavior of their statistics when they are applied for the forecast even though they match the data with high precision. In addition to these moment quantity dynamical terms, forcing terms are also imposed on the model identification. These forcing terms do not depend on $m$, $v$, $s$, or $k$, but they can be functions of time $t$. To be consistent with the physical justifications in Section \ref{Subsec:Setup}, the forcing is divided into three periods. Linear functions of $t$ are adopted for the first and second periods, which are the natural choices corresponding to the linear driving stresses. On the other hand, the linear driving stress interacting with the underlying nonlinear dynamics induces a strong nonlinear response of the time evolution of the moments associated with the von Mises stress in the third period, which cannot be fully characterized by adopting a linear forcing. To this end, a quadratic function of $t$ is utilized for this last period. Different forcing terms are imposed on each moment equation as a starting point,
\begin{equation}
F(t) = \left\{\begin{array}{l}
f_1 + f_2 t,\quad t_0 < t \leq t_1\\
f_3 + f_4 t,\quad t_1 < t \leq t_2 \\
f_5 + f_6 t + f_7 t^2,\quad t_2 < t \leq t_3.
\end{array}\right.
\label{forcing}
\end{equation}
The results with various simplifications of the forcing representations will be presented at the end of this subsection.
To determine the model structure, the causation entropy is computed from each candidate function to each of the moments. The threshold value is to be $r=0.09$. If the causation entropy is above this value, then the corresponding candidate function is retained. With these choices of the hyper-parameters, the identified model structure is shown in Figure \ref{Model_Strucuture_Case1} with the associated model parameters listed in Table \ref{Table:Case1}. 
Note that this threshold must be chosen to be positive, as the numerically calculated causation entropy will never be precisely zero. Nevertheless, this threshold should remain small to consider all essential terms in the dynamics. Changing this threshold value will increase or decrease the number of terms in the identified model. But perturbing the threshold within a reasonable range around $r=0.09$ still leads to similar dynamical behavior as the model presented below, indicating the relative robustness of the method.

\begin{table}[th]
\centering
\small
\begin{tabular}{|c|c|c|c|c|c|c|}
\hline
$\theta_{v}^{m}$ & $\theta_{s}^{m}$ & $\theta_{ms}^{m}$ & $\theta_{vk}^{m}$ & $\theta_{sk}^{m}$& $\theta_{kk}^{m}$ & $f_{1}^{m}$\\
\hline
-48.4939 & 7.4860 & -0.1678 & 4.8556 & 3.1938 & 0.6583 & 34.8863\\

\hline
 $f_{2}^{m}$ & $f_{3}^{m}$ & $f_{4}^{m}$ & $f_{5}^{m}$ & $f_{6}^{m}$ & $f_{7}^{m}$ & $\theta_{v}^{v}$ \\
\hline
1.4566 & 34.1355 & -11.8488 & 398.9180 & -608.6578 & 226.6429 & -16.4172 \\	

\hline
$\theta_{s}^{v}$& $\theta_{k}^{v}$ & $\theta_{vs}^{v}$ & $\theta_{kk}^{v}$ & $f_{1}^{v}$ & $f_{2}^{v}$ & $f_{3}^{v}$\\
\hline
-0.8311 & -21.6732 & 27.4578 & 1.9840 & 49.6498 & 52.397 & 63.4550\\

\hline
$f_{4}^{v}$ & $f_{5}^{v}$ & $f_{6}^{v}$ & $f_{7}^{v}$ & $\theta_{m}^{s}$ & $\theta_{mm}^{s}$ & $\theta_{vv}^{s}$\\
\hline
-3.7120 & 188.8415 & -206.3912 & 78.6198 & -11.7769 & 0.1678 & -27.4578\\

\hline
$f_{1}^{s}$ & $f_{2}^{s}$ & $f_{3}^{s}$ & $f_{4}^{s}$ & $f_{5}^{s}$ & $f_{6}^{s}$ & $f_{7}^{s}$\\
\hline
-0.1485 & 405.4597 & 112.4508 & -37.7338 & 1228.0147 & -1750.9335 & 642.2245 \\

\hline
$\theta_{m}^{k}$ & $\theta_{v}^{k}$ & $\theta_{k}^{k}$ & $\theta_{mv}^{k}$ & $\theta_{ms}^{k}$ & $\theta_{mk}^{k}$ & $\theta_{vk}^{k}$ \\
\hline
-2.4556 & 47.0266 & 1.8248 & -4.8556 & -3.1938 & -0.6583 & -1.9840\\

\hline
$f_{1}^{k}$ & $f_{2}^{k}$ & $f_{3}^{k}$ & $f_{4}^{k}$ & $f_{5}^{k}$ & $f_{6}^{k}$ & $f_{7}^{k}$\\
\hline
-5.7400 & 88.5596 & 18.8439 & -15.0779 & 321.0874 & -516.7953 & 208.7485\\
\hline
\end{tabular}
\caption{The parameters in the identified model based on the model structure in Figure \ref{Model_Strucuture_Case2}. Here $\theta_b^a$ means the parameter appearing in the equation of $a$ and is the coefficient of the $b$ term, and $f_i^a$ represents the $i$th coefficient of forcing terms in the equation of $a$.}\label{Table:Case1}
\end{table}

%{\color{red}{Yinling, add a figure as we discussed and add some briefly discussions for the sparsity, the selection criterion, and the interactions between different components, etc.}}

\begin{figure}[ht]\centering
\hspace*{-0cm}\includegraphics[width=1.0\textwidth]{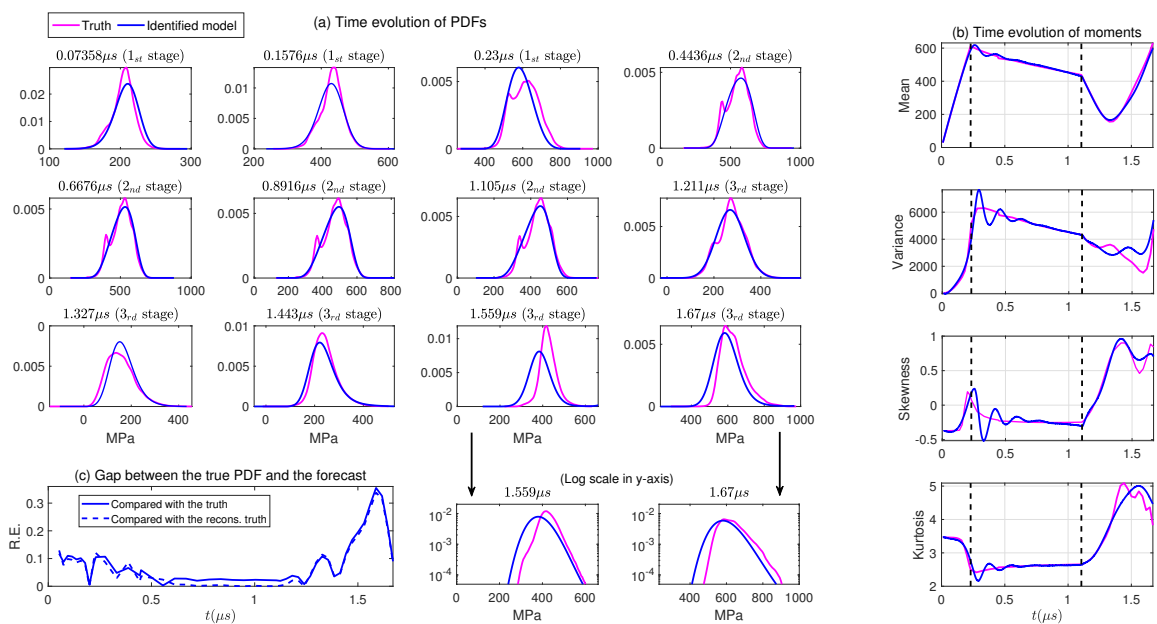}
\caption{Experiment I: Comparison of the truth (pink) and the predicted statistics using the moment equations (blue) for simulation \#1. Panel (a): time evolution of the PDFs. Panel (b): time evolution of the leading four moments. Panel (c): time evolution of the relative entropy (R.E.) that quantifies the difference between the truth (solid curve) or the reconstructed truth based on the true values of the leading four moments (dashed curve) and the predicted PDF using the moment equations. }\label{Identified_Case1}
\end{figure}

Figure \ref{Identified_Case1} compares the truth (pink) and the predicted statistics from the identified model (blue) starting from $t=0$ for simulation \#1. The time evolution of the moments is predicted quite accurately, especially for the first and second time periods. The details of the complicated structures in the third time period are not entirely captured. But the overall tendency is predicted by the identified moment equations. First, the predicted time evolution of the mean from the model is nearly identical to the truth, which indicates that the statistical reduced-order model well captures the averaged behavior of the sophisticated physical model. Second, the model perfectly describes the kurtosis being slightly larger than 3 (the Gaussian value) during the first compression stage and decreasing to around 2.5 during the second stage with the compressive stress. In other words, the precursor of the extreme damage events is appropriately characterized by the model. The model prediction also follows the rapid increase of the kurtosis in the third stage and provides an accurate forecast of the strong positive skewness. Capturing these two significant features allows the model to precisely predict the one-sided fat tail distribution characterizing the uncertainty resulting from extreme events.
Based on the skillful recovery of the moments, the reconstructed PDFs also illustrate similar features as the truth in most instances. The only exception to the accurate PDF forecast occurs around $t=1.559\mu s$ due to a slight overestimation of the variance. The error in the forecast PDF around that time instant mainly comes from the inaccurate estimation of probability for moderate or small stress values. Nevertheless, it is worth highlighting that the tail behavior on the positive side for the time instants is consistently well predicted, including the one at $t=1.559\mu s$ (see the panels with the y-axis showing in the logarithm scale). This is a significant feature, which indicates that the uncertainty quantification of the extreme events using the statistical reduced-order model is entirely accurate throughout time. In other words, the model can provide a precise probabilistic forecast of the timing of the occurrence of extreme damage events, similar to the one computed from the much more expensive and sophisticated physical model.

\begin{figure}[ht]\centering
\hspace*{-0cm}\includegraphics[width=1.0\textwidth]{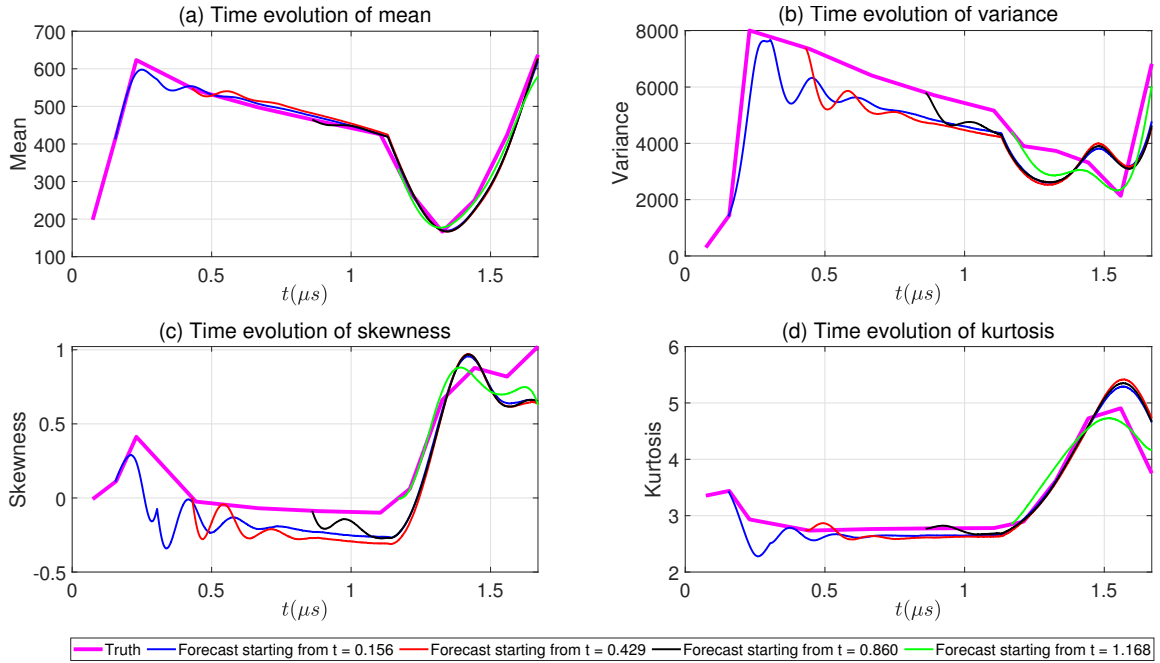}
\caption{Extrapolation test of Experiment I: Applying the identified moment equation from the training realization (simulation \#1) to predict the time evolution of the statistics associated with a different polycrystal realization (simulation \#10; thick pink color). Different thin curves show the forecast starting from different initial values at different points in time. The exact values of the leading four moments from simulation \#10 serve as the initial values for the forecast using the identified moment equations. }\label{Sensitivity}
\end{figure}

\begin{figure}[ht]\centering
\hspace*{-0cm}\includegraphics[width=1.0\textwidth]{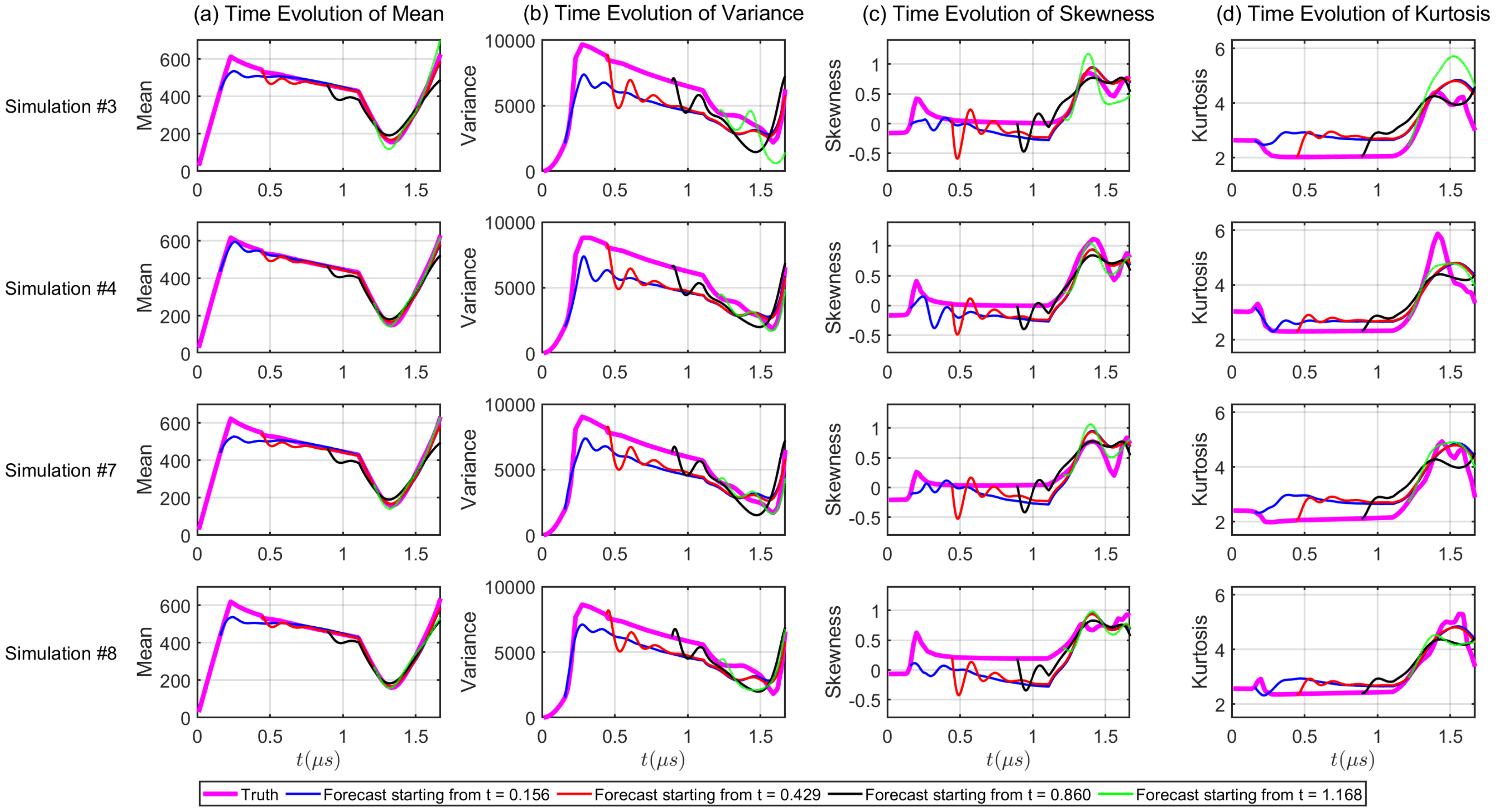}
\caption{Extrapolation tests of Experiment I: Applying the identified moment equation from the training realization (simulation \#1) to predict the time evolution of the statistics associated with four different polycrystal realizations (simulation \#3, simulation \#4, simulation \#7, and simulation \#8). Different thin curves show the forecast starting from different initial values at different points in time. The exact values of the leading four moments from these four different simulations serve as the initial values for the forecast using the identified moment equations.}\label{verification_test4}
\end{figure}

Figure \ref{Sensitivity} shows the results of the extrapolation test. Recall that the moment equations were identified using the first SVE von Mises stress data set (simulation \#1). Extrapolation means such identified moment equations are applied to predict the time evolution of the statistics of the other SVE data sets. Skillful extrapolations indicate that the identified model is generalizable and can be used for efficient forecast beyond the situation of the training data set. The forecast of the identified model for the second SVE data set (simulation \#10; thick pink color) is presented in Figure \ref{Sensitivity}, which results from using the same loading condition but different numbers of grains (in total 66) and microstructure relative to simulation \#1 and shown in Figure \ref{Poly_Meshes}. In these extrapolation tests, the exact values of the leading four moments from simulation \#10 serve as the initial conditions for the forecast using the moment equations identified from the simulation \#1. Simulation \#10 is also used to measure forecast accuracy. It is seen that the forecasts starting from different time instants covering the three stages are all reasonably accurate. Notably, the predicted kurtosis coincides with the truth, which remains at small values below 3 over the second time period and experiences a rapid increase over the third time period. It implies that the predictions of the tail probability that corresponds to the quantification of uncertainty in extreme damage events and its statistical precursors are both accurate. The extrapolation tests for the other four SVE data sets are quantitatively similar and are included in Figure \ref{verification_test4}. The results here indicate that the identified model is skillful and robust for the extrapolation test of predicting different polycrystal realizations.

\begin{figure}[ht]\centering
\hspace*{-0cm}\includegraphics[width=1.0\textwidth]{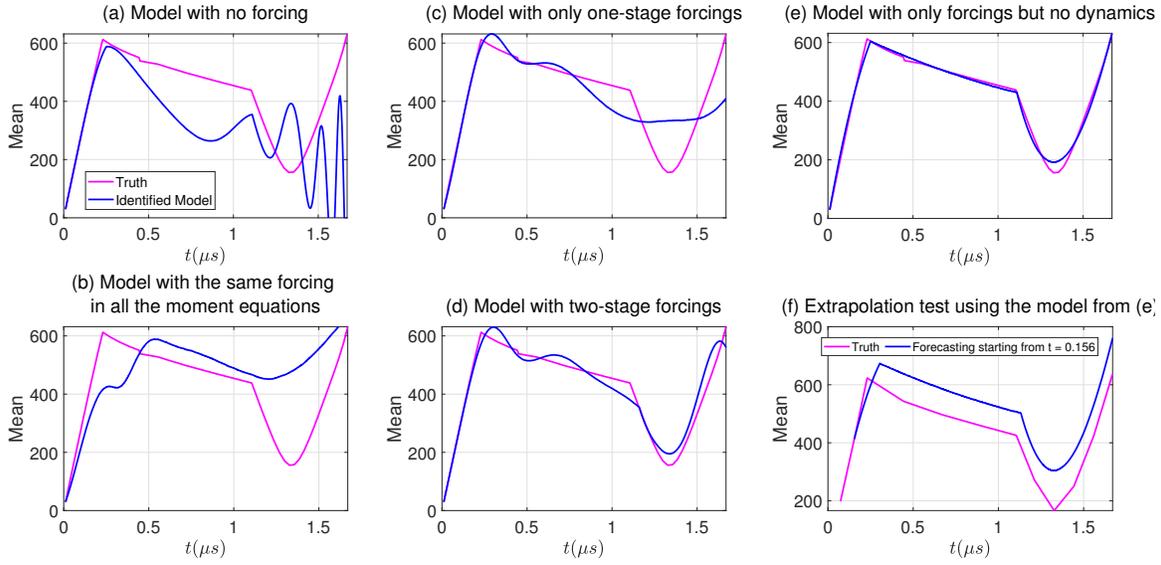}
\caption{Sensitivity tests of model identification in Experiment I. The forecast of the mean using the identified moment equations in different setups. Panel (a): the model without additional forcing. Panel (b): the model with three-stage forcings but the same forcing coefficients are utilized in all the four moment equations. Panel (c): the model with only a one-stage forcing that is a quadratic function of time. Panel (d): the model with only two-stage forcings that are a linear and a quadratic function of time, respectively. The first stage is from $t=0$ to $t=1.105\mu s$ while the second stage is from $t=1.015\mu s$ to $1.67\mu s$. Panel (e): the model without dynamics but only forcing. Panel (f): the extrapolation test to a different realization using the identified model from Panel (e).   }\label{Models_Different_Forcing}
\end{figure}

Finally, Figure \ref{Models_Different_Forcing} includes additional tests of the model identification. Similar to the previous setup, the model is learned from simulation \#1, while the extrapolation test is based on simulation \#10. Only the time evolution of the mean and its forecast using different identified models is shown. The qualitative results remain the same for other moments, and, in general, higher-order moments have more significant errors. The result in Panel (a) reveals that the forcing terms play an important role in capturing the time evolution of the moments. Without the forcing terms, the optimal identified model built upon the same library is insufficient in characterizing the transitions in the time evolution of the moments. Adding more candidate functions and increasing the complexity of the model can improve the fitting results but will make the model lose the extrapolation skill. Panel (b) shows the forecast using the identified model, where the same forcing terms and coefficients are applied to all four moment equations. This critical test indicates that different moments driven by the same external forcing in an additive way fail to provide a skillful model. This suggests that if there is one common external forcing, it must interact with different moments in a complicated nonlinear way. This conjuncture will be further validated in Section \ref{Subsec:Experiment_II}. Panel (c) illustrates that if the model is imposed with only a one-stage forcing that is a quadratic function of time, then the model is again insufficient to characterize the time evolution of the moments. Nevertheless, Panel (d) suggests that a model with two-stage forcings that combines the compression and the tension stages can provide good results. Here, the two forcings are a linear and a quadratic function of time, respectively. In addition to the mean, the forecast of higher-order moments and the extrapolation have comparable high skills. The error is slightly larger than the model with a three-stage forcing, but the model shown in Panel (d) has a somewhat simpler form, which can be an appropriate model. Finally, Panels (e)--(f) imply that although the identified model with only the forcing but no other dynamics can provide good fitting results, the initial error in the extrapolation test will never be suppressed. Such an undesirable feature in the extrapolation test is due to the lack of damping effect in the identified model, which is crucial to mitigate the initial or system biases. It also suggests that the identified model should not be driven entirely by forcing terms.

\subsection{Experiment II: Learning the moment equations of variance, skewness and kurtosis driven by the mean forcing}\label{Subsec:Experiment_II}
Recall that many existing macroscale ductile damage models representing the formation and evolution of porosity fields under dynamic loading employ mean-valued (first moment) material internal state variables to characterize the evolution of stress, strain, material structure, and porosity with deformation over time. Similarly, the time evolution of the mean state variables can also be measured relatively accurately from lab experiments based on a small number of sample points. Therefore, it is natural to treat the mean as a known component and use such information to identify the high-order moments. This subsection aims to learn a three-dimensional moment equation for the time evolution of variance, skewness, and kurtosis, where the mean is treated to be known. In addition to this practical reasoning, another crucial motivation for this experiment is to investigate the relationship between the mean and the high-order moments via the learning process. Particularly, it is essential to reveal how the mean interacts with the high-order moments in determining the moment equation of the latter. As the information of extreme events lies in these high-order moments, the contribution from the mean to quantifying uncertainty in extreme events is essential to understand.

\begin{figure}[th]\centering
\hspace*{-0cm}\includegraphics[width=1.0\textwidth]{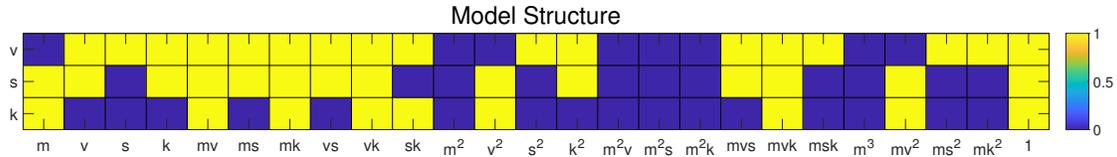}
\caption{The identified model structure of the statistical reduced-order model in Experiment II. The three rows show the identified structure on the right-hand side of the equation of variance ($v$), skewness ($s$), and kurtosis ($k$), respectively. Note that the mean ($u$) is known in this model. The terms represented by the yellow color are the ones that exist in the identified model while those indicated by the blue color do not appear in the model. }%{\color{red}{Yinling, replace this figure by the correct one.}}  }
\label{Model_Strucuture_Case2}
\end{figure}

\begin{table}[th]
\centering
\small
\begin{tabular}{|c|c|c|c|c|c|c|c|c|c|}
\hline
$\theta_{v}^{v}$ & $\theta_{s}^{v}$ & $\theta_{k}^{v}$ & $\theta_{mv}^{v}$ & $\theta_{ms}^{v}$ & $\theta_{mk}^{v}$& $\theta_{vs}^{v}$ & $\theta_{vk}^{v}$ & $\theta_{sk}^{v}$ & $\theta_{ss}^{v}$ \\
\hline
135.5376 & -16.237 & -51.894 & -22.5021 & 16.4371 & 3.4971 & 96.6142 & -48.5493 & 2.1524 & -26.3572\\
\hline
$\theta_{kk}^{v}$ & $\theta_{mvs}^{v}$ & $\theta_{mvk}^{v}$ & $\theta_{msk}^{v}$ & $\theta_{mss}^{v}$ & $\theta_{mkk}^{v}$& $\theta_{1}^{v}$ & $\theta_{m}^{s}$ & $\theta_{v}^{s}$ & $\theta_{k}^{s}$ \\
\hline
9.0825 & -15.7718 & 7.1996 & -3.9981 & 8.7155 & -0.8878 & 70.0347 & 2.6882 & 64.5419 & 3.5754 \\
\hline
$\theta_{mv}^{s}$ & $\theta_{ms}^{s}$ & $\theta_{mk}^{s}$ & $\theta_{vs}^{s}$ & $\theta_{vk}^{s}$ & $\theta_{vv}^{s}$ & $\theta_{kk}^{s}$ & $\theta_{mvs}^{s}$ & $\theta_{mvk}^{s}$ & $\theta_{mvv}^{s}$\\
\hline
-23.6133 & 1.7053 & -0.309 & 26.3572 & -2.1524 & -96.6142 & -1.6809 & -8.7155 & 3.9981 & 15.7718\\
\hline
 $\theta_{1}^{s}$ & $\theta_{m}^{k}$ & $\theta_{mv}^{k}$ & $\theta_{mk}^{k}$ & $\theta_{vk}^{k}$ & $\theta_{sk}^{k}$ & $\theta_{vv}^{k}$& $\theta_{mvk}^{k}$& $\theta_{mvv}^{k}$& $\theta_{1}^{k}$\\
\hline
4.8621 & -3.4606 & 3.4666 & -0.2942 & -9.0825 & 1.6809 & 48.5493 & 0.8878 & -7.1996 & 15.3728\\
\hline
\end{tabular}\caption{The parameters in the identified model based on the model structure in \ref{Model_Strucuture_Case2}. Here $\theta_b^a$ means the parameter appearing in the equation of $a$ and is the coefficient of the $b$ term.}\label{Table:Case2}
\end{table}

Figure \ref{Model_Strucuture_Case2} shows the identified model structure using simulation \# 1 results as in Experiment I .  To balance model sparsity and essential dynamics, the threshold values for the three moments (variance, skewness and kurtosis) are set to be $r_v = 0.01,\ r_s = 0.04$, and $r_k = 0.12$, respectively. Different from experiment I, the model here does not require to have additional time-dependent forcing. This leads to the first conclusion that the high-order moments are driven by the mean. Next, it is worthwhile to highlight that
the mean does not play a role as a simple additive forcing. Instead, the mean interacts with all the other moments in a highly nonlinear way through multiplicative terms, for example, $mv$ and $mk$. In other words, knowing the time evolution of the mean significantly facilitates the forecast uncertainty of extreme events. This is consistent with the conclusion from Panel (b) of Figure \ref{Models_Different_Forcing} that indicates the strong nonlinear nature of the moment equations that reflect the complicated nonlinear dynamics throughout the SVE deformation process.

\begin{figure}[ht]\centering
\hspace*{-0cm}\includegraphics[width=1.0\textwidth]{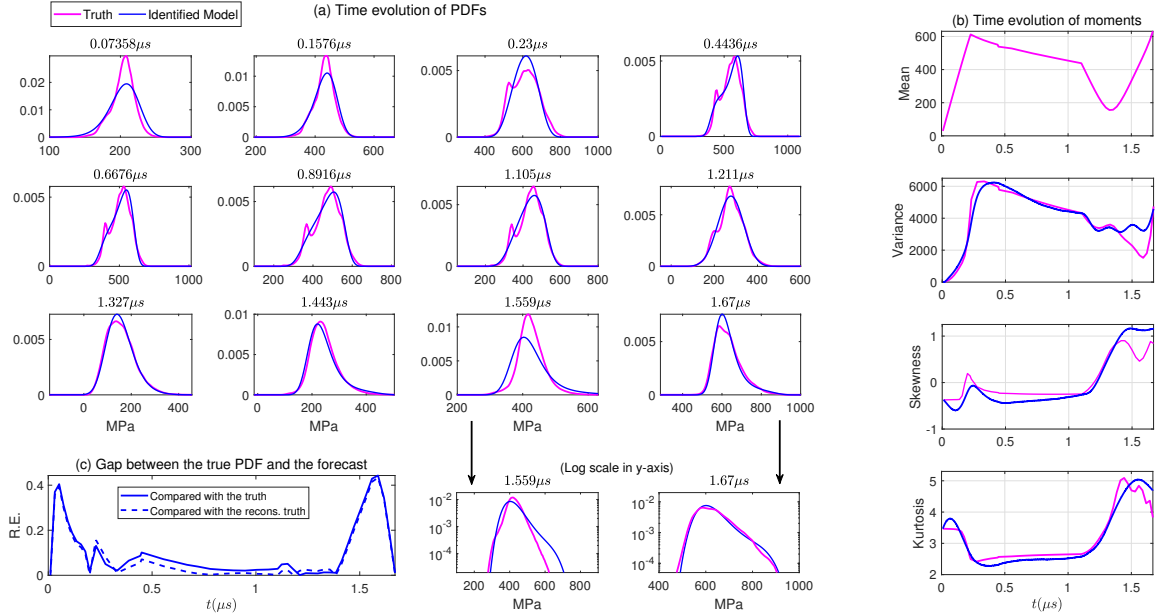}
\caption{Experiment II: Comparison of the truth (pink) and the predicted statistics using the moment equations (blue). Panel (a): time evolution of the PDFs. Panel (b): time evolution of the leading four moments. The first moment (mean) is known in the identified model. Panel (c): time evolution of the relative entropy (R.E.) that quantifies the difference between the truth (solid curve) or the reconstructed truth based on the true values of the leading four moments (dashed curve) and the predicted PDF using the moment equations. }
\label{Identified_Case2}
\end{figure}

\begin{figure}[ht]\centering
\hspace*{-0cm}\includegraphics[width=1.0\textwidth]{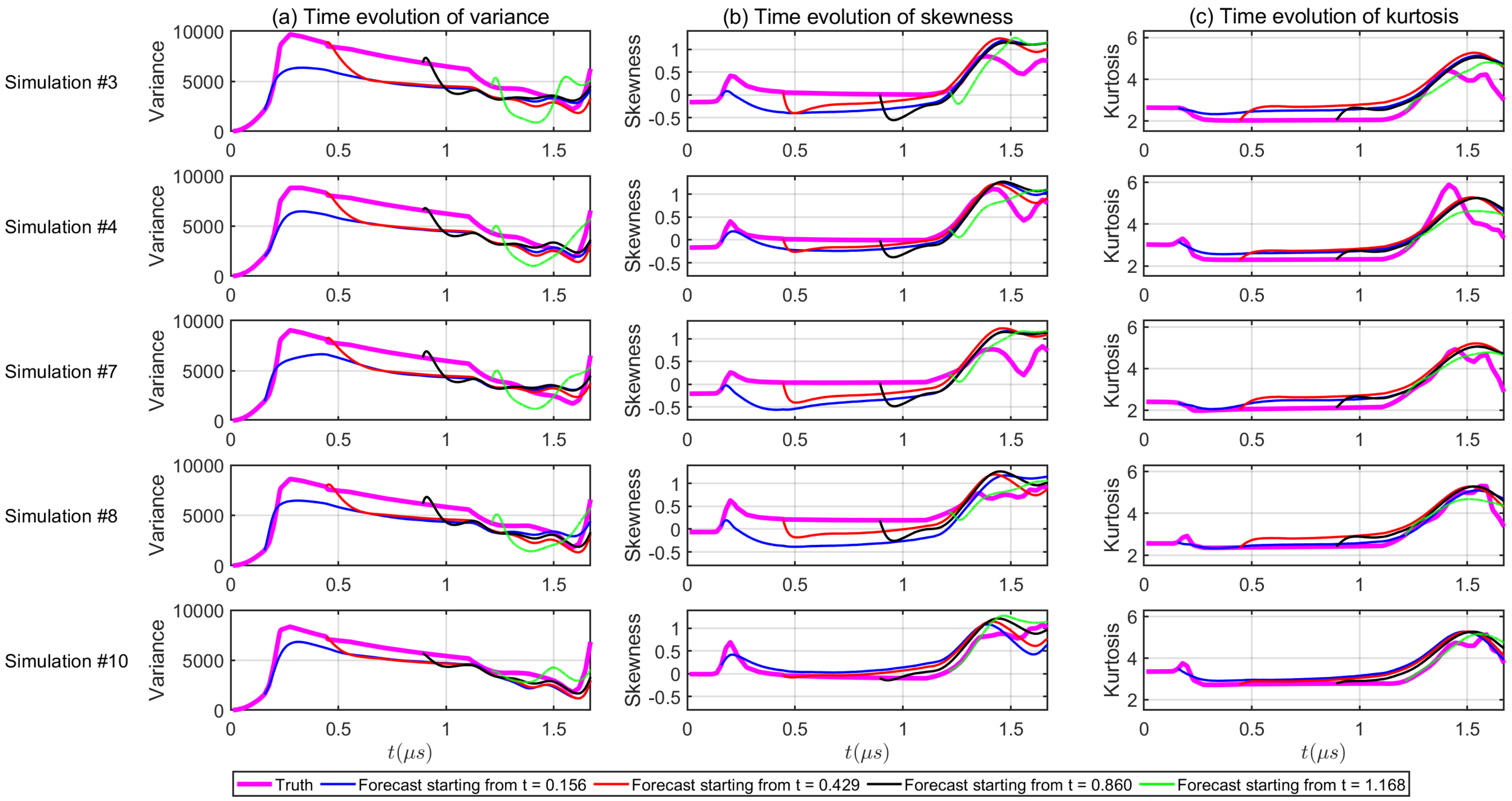}
\caption{Extrapolation tests of Experiment II: Applying the identified moment equation from the training realization (simulation \#1) to predict the time evolution of the statistics associated with five different polycrystal realizations (simulation \#3, simulation \#4, simulation \#7, simulation \#8, and simulation \#10). Different thin curves show the forecast starting from different initial values. The exact values of the leading three moments from these five simulations serve as the initial values for the forecast using the identified moment equations.}%{\color{red}{Yinling, replace this figure by the correct one.}}    }
\label{verification_test4_m}
\end{figure}

Figure \ref{Identified_Case2} includes the time evolution of the statistics from the model identification. The results are similar to those in Figure \ref{Identified_Case1}. The identified model is overall accurate except at the time instants around $t = 1.559\mu s$, where the variance is slightly overestimated. Nevertheless, the probability of the extreme events corresponding to the fat tail on the positive side is overall captured reasonably accurate over time, especially given the fact that in the absence of time-dependent forcing the identified model structure here is simpler than that in Experiment I. This implies an overall accurate uncertainty quantification of extreme events. This means the identification model can be applied to the statistical forecast of extreme damage events that provide the timing and the probability of the possible occurrence of the ductile damage. Figure \ref{verification_test4_m} shows the extrapolation results by applying the identified model to the other five polycrystal realizations. The predicted moments starting from different time instants follow the truth reasonably well, which indicates the skillful generalization of the model.

\section{Discussion and Conclusion}\label{Sec:Conclusion}

In this paper, a new data-driven reduced-order modeling framework is developed and applied to study the early-stage process of porosity-based ductile damage in polycrystalline metallic materials. The framework starts with computing the time evolution of the leading few moments of specific state variables from the spatiotemporal solution of the polycrystal simulation. Then a simple causality-based sparse model identification algorithm, including essential physical constraints, is utilized to discover the governing equations of these moments. An approximate solution of the time evolution of the PDF is obtained from the predicted moments and the maximum entropy principle. Numerical experiments based on polycrystal realizations of a representative BCC tantalum illustrate a skillful reduced-order model in characterizing the time evolution of the non-Gaussian PDF of the von Mises stress and quantifying the probability of extreme events. The learning process also reveals that the mean stress is not simply an additive forcing to drive the higher-order moments and extreme events. Instead, it interacts with the higher-order moments in a strongly nonlinear and multiplicative fashion. Finally, the calibrated moment equations provide a reasonably accurate forecast when applied to the realizations outside the training data set, indicating the robustness of the model and the skill for extrapolation.

These results also demonstrate the importance of including measures of material statistics of the higher moments into probabilistic descriptions of porosity-based ductile damage. The polycrystal simulations indicate noticeable structural evolution taking place during the shock compression loading as indicated by significant magnitude of von Mises stress during the stage of unloading after passage of the shock wave through the center of the sample. This is followed by high tensile reverse loading during which significantly high values of skewness and kurtosis were observed. In particular, kurtosis values reach magnitudes of 5 in this region of tensile loading which indicates fat tails of the stress distribution which deviate far from the mean stress state and large magnitudes. It is expected that the regions of the body experiencing such particularly high stress states to also be more probable to nucleate damage events. Such fat tails as quantified by large values of kurtosis are important as this segment of loading also immediately precedes damage events observed experimentally and computationally via the macroscale damage model. Although small in volume, the SVEs are statistically related to the material used experimentally and are therefore quantitatively relevant. This recommends the possibility of adopted use of reduced-order statistical models as described here to accurately represent the complex structural evolution of materials in a tractable fashion.

One central subject in practice is efficiently forecasting extreme stress/damage events under different loading conditions. Instead of repeatedly running the expensive polycrystal models with different conditions, the statistical reduced-order model allows an alternative way to rapidly reach the associated statistical forecast with appropriate uncertainty quantification. Here, a small number of the statistical reduced-order models can be trained by considering several fundamental loading conditions, such as imposing compression stress, shearing stress, and tension stress. A combination of these fundamental loading states is usually sufficient to represent more complex and arbitrary loading conditions. This means a large number of the statistical forecast of the damage events with various loading conditions can be easily carried out by running these calibrated statistical low-order models, which are computationally much more efficient than running the physical model multiple times. This remains a natural immediate future work.

It is also worth noticing that the highly efficient statistical reduced-order model developed here does not directly tell the exact location of the high-stress events in polycrystalline metallic materials. Nevertheless, the time evolution of the probability of extreme stress values computed from the statistical reduced-order model provides a valuable indicator to predict the timing and the associated uncertainty of the occurrence of extreme damage events with an almost negligible computational cost. The result can be combined with specific spatial distribution maps from other approaches to infer more detailed information about ductile damage.

Finally, the statistical reduced-order model can also be developed based on experimental data. The resulting model is then quite useful for dynamically correcting the error in forecasting the ductile damage from more complicated physical models. This can be achieved by running the statistical reduced-order model and the physical model simultaneously. At each time instant, compute the leading few moments from the spatial solution of the physical model. Correct the error in the distribution associated with these data in light of the output from the statistical reduced-order model. Then redistribute the solution values of the physical model and run it forward up to the next instant. This process exploits the statistical reduced-order model as a bridge to connect experimental data with the physical model to perform efficient data assimilation for model error correction in a dynamic fashion.

\section*{Acknowledgement}
The authors gratefully acknowledge funding for this work from National Science Foundation DMREF program (grant number: \#2118399). CAB also acknowledges support from the DOE Advanced Simulation and Computing program at Los Alamos National Laboratory. HC acknowledges the support from National Research Foundation of Korea (grant numbers: 2020R1C1C101324813 and 2021R1A4A103278312).
\bibliography{references}

\end{document}